\documentclass[preprint, astrosymb]{aastex631}

\newcommand{\Msun}{M$_{\odot}$} 

\shorttitle{SMWLV Star Clusters Roadmap}
\shortauthors{SMWLV - Star Clusters Science Working Group}

\begin{document}
\title{Rubin Observatory LSST Stars Milky Way and Local Volume Star Clusters Roadmap}

\author[0000-0002-7383-7106]{Christopher Usher}
\affiliation{The Oskar Klein Centre, Department of Astronomy, Stockholm University, AlbaNova, SE-10691 Stockholm, Sweden}

\author[0000-0002-8532-4025]{Kristen C. Dage}
\affiliation{Department of Physics, McGill University, 3600 University Street, Montr\'eal, QC H3A 2T8, Canada}

\author[0000-0002-6301-3269]{L\'eo Girardi}
\affiliation{INAF -- Osservatorio Astronomico di Padova, Vicolo dell'Osservatorio 5, I-35122 Padova, Italy}

\author[0000-0003-2767-0090]{Pauline Barmby} \affiliation{Dept of Physics \& Astronomy and Institute for Earth \& Space Exploration, University of Western Ontario, London, Canada}

\author[0000-0002-4102-1705]{Charles J.	Bonatto} \affiliation{Instituto de Física, Universidade Federal do Rio Grande do Sul (UFRGS), Porto Alegre, RS, Brazil}\affiliation{Laboratório Interinstitucional de e-Astronomia - LineA, Rua Gal. José Cristino 77, Rio de Janeiro, RJ - 20921-400, Brazil}

\author[0000-0003-3220-0165]{Ana L. Chies-Santos} \affiliation{Instituto de Física, Universidade Federal do Rio Grande do Sul (UFRGS), Porto Alegre, RS, Brazil}\affiliation{Laboratório Interinstitucional de e-Astronomia - LineA, Rua Gal. José Cristino 77, Rio de Janeiro, RJ - 20921-400, Brazil}

\author[0000-0002-2577-8885]{William I.	Clarkson} \affiliation{Natural Sciences Department, University of Michigan-Dearborn, 4901 Evergreen Road, Dearborn, MI 48128, USA}

\author[0000-0002-4430-9427]{Matias	G\'omez Camus} \affiliation{Instituto de Astrofisica, Universidad Andres Bello, Fernandez Concha 700, Las Condes, Santiago, Chile}

\author[0000-0002-2527-8223]{Eduardo A.	Hartmann} \affiliation{Instituto de Física, Universidade Federal do Rio Grande do Sul (UFRGS), Porto Alegre, RS, Brazil}
\affiliation{Laboratório Interinstitucional de e-Astronomia - LineA, Rua Gal. José Cristino 77, Rio de Janeiro, RJ - 20921-400, Brazil}

\author[0000-0001-7934-1278]{Annette  M. N. Ferguson} \affiliation{ Institute for Astronomy, University of Edinburgh Royal Observatory, Blackford Hill, Edinburgh EH9 3HJ, UK}

\author[0000-0001-9186-6042]{Adriano	Pieres} \affiliation{Laboratório Interinstitucional de e-Astronomia - LineA, Rua Gal. José Cristino 77, Rio de Janeiro, RJ - 20921-400, Brazil}

\author[0000-0002-8893-2210]{Loredana	Prisinzano} \affiliation{INAF -- Osservatorio Astronomico di Palermo, Piazza del Parlamento, 1 90129 Palermo}

\author[0000-0001-8283-4591]{Katherine L.	Rhode} \affiliation{Department of Astronomy, Indiana University, Bloomington, IN 47405, USA}
\author{R. Michael 	Rich} \affiliation{Department of Physics and Astronomy, UCLA, PAB 430 Portola Plaza, Los Angeles, CA 90095-1547}
\author[0000-0003-1801-426X]{Vincenzo	Ripepi} \affiliation{INAF-Osservatorio Astronomico di Capodimonte, Salita Moiariello 16, 80131, Napoli, Italy}

\author{Basilio Santiago} \affiliation{Instituto de Física, Universidade Federal do Rio Grande do Sul (UFRGS), Porto Alegre, RS, Brazil}
\affiliation{Laboratório Interinstitucional de e-Astronomia - LineA, Rua Gal. José Cristino 77, Rio de Janeiro, RJ - 20921-400, Brazil}

\author[0000-0002-3481-9052]{Keivan G.	Stassun} \affiliation{Department of Physics and Astronomy, Vanderbilt University, Nashville, TN 37235, USA}

\author[0000-0001-6279-0552]{R.A.	Street} \affiliation{Las Cumbres Observatory, 6740 Cortona Drive, Suite 102, Goleta, CA 93117, USA}
\author[0000-0002-3258-1909]{R\'obert	Szab\'o} \affiliation{Konkoly Observatory, CSFK, MTA Centre of Excellence, H-1121 Konkoly Thege Mikl\'os \'ut 15-17, Budapest, Hungary}
\affiliation{MTA CSFK Lend\"ulet Near-Field Cosmology Research Group H-1121 Konkoly Thege Mikl\'os \'ut 15-17, Budapest, Hungary}

\author[0000-0002-4115-0318]{Laura	Venuti} \affiliation{SETI Institute, 339 Bernardo Ave, Suite 200, Mountain View, CA 94043, USA}

\author[0000-0001-6081-379X]{Simone	Zaggia} \affiliation{INAF -- Osservatorio Astronomico di Padova, Vicolo dell'Osservatorio 5, I-35122 Padova, Italy}

===

\author[0000-0002-1261-1015]{Marco  Canossa} \affiliation{Instituto de Física, Universidade Federal do Rio Grande do Sul (UFRGS), Porto Alegre, RS, Brazil}\affiliation{Laboratório Interinstitucional de e-Astronomia - LineA, Rua Gal. José Cristino 77, Rio de Janeiro, RJ - 20921-400, Brazil}

\author[0009-0008-4034-7670]{Pedro Floriano} \affiliation{Instituto de Física, Universidade Federal do Rio Grande do Sul (UFRGS), Porto Alegre, RS, Brazil}\affiliation{Laboratório Interinstitucional de e-Astronomia - LineA, Rua Gal. José Cristino 77, Rio de Janeiro, RJ - 20921-400, Brazil}

\author[0009-0005-7299-4168]{Pedro  Lopes} \affiliation{Instituto de Física, Universidade Federal do Rio Grande do Sul (UFRGS), Porto Alegre, RS, Brazil}
\affiliation{Laboratório Interinstitucional de e-Astronomia - LineA, Rua Gal. José Cristino 77, Rio de Janeiro, RJ - 20921-400, Brazil}

\author[0009-0009-3088-5886]{Nicole L.  Miranda} \affiliation{Instituto de Física, Universidade Federal do Rio Grande do Sul (UFRGS), Porto Alegre, RS, Brazil}
\affiliation{Laboratório Interinstitucional de e-Astronomia - LineA, Rua Gal. José Cristino 77, Rio de Janeiro, RJ - 20921-400, Brazil}

\author[0000-0002-4778-9243]{Raphael A. P.  Oliveira} \affiliation{Universidade de São Paulo, IAG, Rua do Matão 1226, São Paulo 05508-090, Brazil}
\affiliation{Laboratório Interinstitucional de e-Astronomia - LineA, Rua Gal. José Cristino 77, Rio de Janeiro, RJ - 20921-400, Brazil}

\author[0000-0002-8556-4280]{Marta  Reina-Campos} \affiliation{Department of Physics \& Astronomy, McMaster University, 1280 Main Street West, Hamilton, L8S 4M1, Canada}
 \affiliation{Canadian Institute for Theoretical Astrophysics (CITA), University of Toronto, 60 St George St, Toronto, M5S 3H8, Canada}
 \author[0000-0002-1379-4204]{A.    Roman-Lopes} \affiliation{Department of Astronomy - Universidad de La Serena}
\author[0000-0002-4989-0353]{Jennifer   Sobeck} \affiliation{Canada-France-Hawaii Telescope, 65-1238 Mamalahoa Hwy, Waimea, HI 96743, USA}



\begin{abstract}
   The Vera C. Rubin Observatory will undertake the Legacy Survey of Space and Time, providing an unprecedented, volume-limited catalog of star clusters in the Southern Sky, including Galactic and extragalactic star clusters. The Star Clusters subgroup of the Stars, Milky Way and Local Volume Working Group has identified key areas where Rubin Observatory will enable significant progress in star cluster research. This roadmap represents our science cases and preparation for studies of all kinds of star clusters from the Milky Way out to distances of tens of megaparsecs.
\end{abstract}
 
\tableofcontents

\section{Introduction}
\label{section:intro}
The Vera C. Rubin Observatory is currently under construction on Cerro Pachón in Chile. Over the course of ten years, the Rubin Observatory will carry out the Legacy Survey of Space and Time \citep[LSST,][]{2019ApJ...873..111I}, imaging 18 000 square degrees of the southern sky over 760 times (and a further 9 000 square degrees at least 120 times as per the
\texttt{baseline\_v3.0\_10yrs} Opsim run; \citealt{2014SPIE.9150E..15D}) in six optical filters ($ugrizy$).
With a median $5\sigma$ point source depth of $r = 24$ per visit and median co-added depth of $r = 26.9$ (in the wide fast deep survey area as per the \texttt{baseline\_v3.0\_10yrs} Opsim run) as well as a median effective seeing of 1.0 arcsec, LSST will provide deep, high-quality imaging over an unprecedented area.

Rubin/LSST Science Collaborations are independent federations of worldwide communities of scientists, self organized based on their interests and expertise. They are built primarily to prepare for LSST observations -- e.g., by providing expert advice and analysis to Rubin; by training, educating and engaging the scientific community;  and by carrying out other activities such as fund-raising, software development, and implementing research inclusion practices. The Star Clusters Science Working Group (SCSWG) was recreated in 2022 inside the Stars, Milky Way and Local Volume (SMWLV) collaboration to promote these activities in regards to star clusters. While firmly based on the SMWLV collaboration, this working group has strong connections with the Galaxies and the Transients and Variable Stars (TVS) Science Collaborations.

Star clusters, both within our own galaxy and beyond, are a window onto a wide range of astrophysical processes, including the following questions:
\begin{itemize}
    \item How do stars form, evolve and die?
    
    \item How do star clusters form, evolve and dissolve?
    
    \item How do galaxies form and evolve?
    
    \item What is the dark matter content of galaxies?
    
\end{itemize}

LSST will allow us to observe the nearest open clusters at distances of less than 100 pc and the most massive globular clusters out to a redshift of $z = 0.06$ with the same survey.
Studying both Galactic and extragalactic star clusters in the same survey is complementary in terms of the scientific goals that can be achieved. 
Star clusters in our own Galaxy can be studied in much greater detail than extragalactic ones.
Observations of extragalactic star clusters allow them to be studied in galaxies with a wide range of masses, morphologies and environments. This allows us to understand what features of the Milky Way's star clusters are universal and what varies.

Although this roadmap was prepared by the SMWLV Star Clusters working group, star cluster science extends beyond the SMWLV collaboration to encompass the interests of the Galaxies, Dark Energy and Transients and Variable Stars collaborations.

A bewildering range of terms including open clusters, globular clusters, young massive clusters, super star clusters, faint fuzzies \citep[e.g.][]{2002AJ....124.1410B, 2013A&A...559A..67C, 2023MNRAS.518.3164R}, extended star clusters \citep[e.g.][]{2005MNRAS.360.1007H, 2015MNRAS.447L..40B}, diffuse star clusters  \citep[e.g.][]{2006ApJ...639..838P} and ultra compact dwarfs \citep[e.g.][]{2012A&A...547A..65B} are used to describe different populations of star clusters, with different parts of the community (i.e. Galactic versus extragalactic) using different terms to describe objects with similar physical properties.

The traditional distinction between old, metal-poor, massive and dense globular clusters in the halo and young, metal-rich, low mass and diffuse open clusters in the disk breaks down when we go other galaxies (and even in the Milky Way is partially due to observational biases).
Many modern theories \citep[e.g.][]{1997ApJ...480..235E, 2015MNRAS.454.1658K} of star cluster formation posit a common origin for the diverse population star clusters we observe.

A distinction, however, should be made between non-nuclear star clusters and nuclear star clusters, since instead of being formed in a single burst of star formation, nuclear star clusters are built up through a combination of the in-spiral of massive star clusters via dynamical friction and in-situ star formation over an extended period of time.
A distinction should also be made between non-nuclear star clusters and objects such as $\omega$ Cen and M54, which are believed to be the stripped nuclei of accreted galaxies. We prefer the term candidate nuclear remnant (CNR) for these stripped objects to the more common term of ultra-compact dwarf (UCD), given that the observed population of UCDs contains a mix of massive or extended star clusters and NRs and that there is a lack of consensus about the definition of UCDs.

In this road map, we use the term young cluster to refer to star clusters with ages of a few hundred million years and younger and globular clusters for star clusters older than this. This young/old distinction is mostly between epochs dominated by cluster formation and dissolution processes, but in stellar evolution terms, it roughly corresponds to when the red giant branch starts to make a significant contribution to the star cluster's luminosity. Among the old clusters, we can also define the class of ``intermediate-aged'' clusters, which is useful from the point of view of the theory of stellar populations: such clusters are characterised by the presence of stars with convective cores at the turn-off and an extended red giant branch (RGB). 
We note that most of the observational techniques used to identify and study star clusters, both resolved and unresolved, are similar for young and old star clusters.

\section{Milky Way Star Clusters}
Star clusters in the Milky Way are fundamental for the calibration of stellar evolution and stellar population models, and serve as important probes of the formation history of the Galaxy. Their study has significantly improved since the availability of Gaia proper motions and parallaxes \citep[see][for a review]{cantatgaudin22}, which provided more reliable distances, orbital information, membership probabilities and a number of new candidate clusters. The significantly deeper time-series photometry from LSST will likely allow us to extend the Gaia revolution, with a more complete sampling of the fainter stellar members, and of more distant or extincted star clusters.  By combining Gaia and LSST photometry/astrometry, it will be possible to catalog more distant members and to associate stellar streams with clusters.  Thanks to its high resolution and multicolor photometry, LSST has the prospect of identifying new clusters in the Galactic plane.
In fields too extinct or dense for Gaia, LSST will provide the astrometry that will enable cluster members to be separated from the field.  If deep enough, $u$-band photometry will prove crucial in disentangling multiple populations.


\subsection{Which MW clusters will LSST observe, and to which depth?}
\label{sec:MWclustercensus}


With the optimization of LSST cadence and footprint being well advanced \citep[see][]{bianco22}, now we have a close-to-final idea of which MW clusters will be observed, and to which depth. This, has been evaluated for the baseline plan \verb$baseline_v2.1_10yrs$, in the following way:
\begin{enumerate}
    \item We create a combined list of Galactic globular clusters (GGC) from the \citet{baumgardt21} and \citet{harris96} catalogs. It contains 160 objects.
    \item We eliminate the GGCs from the \citet{kharchenko13} catalog, producing a clean list of open clusters (OCs) with 2862 objects. Although not reflecting the more complete census of MW star clusters provided by the latest Gaia data releases \citep[see e.g.,][]{cantatgaudin18,castroginard20,Kounkel:2020,cantatgaudin22,hao22,hunt23}, \citet{kharchenko13}'s catalog still provides essentially the same distribution in terms of distances and ages.
    \item In the metrics analysis facility \citep[MAF;][]{jones14}, we provide their central coordinates to \texttt{maf.UserPointsSlicer} and evaluate a series of metric values, comprising \texttt{maf.Coaddm5Metric()} and \texttt{maf.CrowdingM5Metric(crowding\_error=0.25)} (hereafter \texttt{CoaddM5} and \texttt{CrowdingM5}, for short), for all 6 LSST filters. 
\end{enumerate}

Already at this stage\footnote{See the current working version of the cluster list at the url  \url{https://drive.google.com/drive/folders/1p1p1btLV4iUJamU1bu_UgFA5aKDePBXB?usp=sharing}. The list will be updated by the SCSWG according to the latest baseline plans and available cluster catalogs.}, 
only clusters inside the footprint$-$totalling 144 GGCs and 2022 OCs$-$provide valid metric values. About 2/3 of these GGCs are found in the main wide fast footprint, the others being in the Galactic Plane minisurvey. For OCs, these fractions reverse, with $\sim1/3$ of the OCs centres being in the WFD footprint. Curiously, there is one GGC in a deep drilling field (namely the remote Arp-Madore 1 cluster, in the Euclid deep drilling field).

The above-mentioned \texttt{CoaddM5} and \texttt{CrowdingM5} metrics represent, respectively, the $5\sigma$ limit for stellar photometric detections, and the limit at which photometric errors induced by crowding (due to the foreground/background MW populations) start causing severe incompleteness (above $\sim50$~\%; see Clarkson et al., in prep.). Therefore, the minimum between these two limits, for every filter $\lambda$, provides the faintest stars to be measured in the clusters:
\begin{equation}
    m_\lambda = \min(\texttt{CoaddM5}_\lambda,\texttt{CrowdingM5}_\lambda) 
    \label{eq:apparent}
\end{equation}
For GGCs, this limit is valid only in the cluster outskirts, since their inner regions will be affected by additional ``self-crowding''. Using the cluster tabulated distances and extinctions, we can also derive the maximum depth in absolute magnitudes, $M_\lambda$:
\begin{equation}
    M_\lambda = m_\lambda - \mu_0 - E_\lambda \times A_V
   \label{eq:absolute}
\end{equation}
where $\mu_0$ is the true distance modulus and $E_\lambda$ is the ratio between the extinction in every filter and the extinction in the $V$ band, $A_V$. $E_\lambda$ values are taken from \citet{chen19}.

\begin{figure*}
    \centering
    \includegraphics[width=0.49\textwidth]{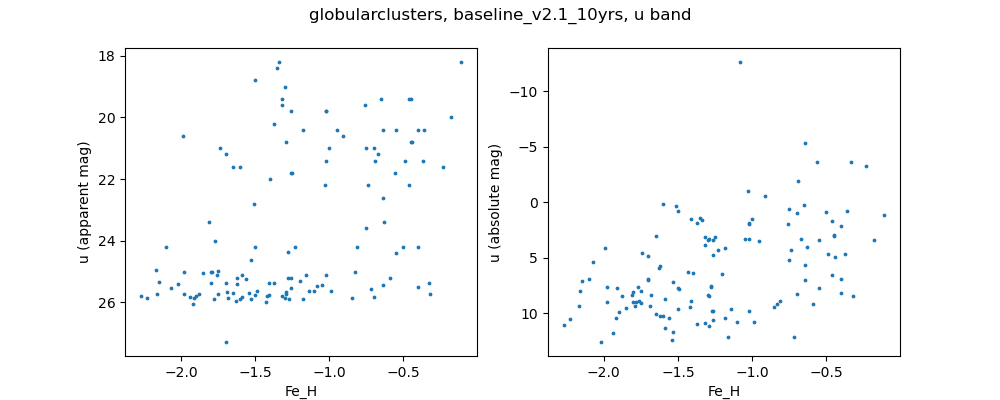}
    \includegraphics[width=0.49\textwidth]{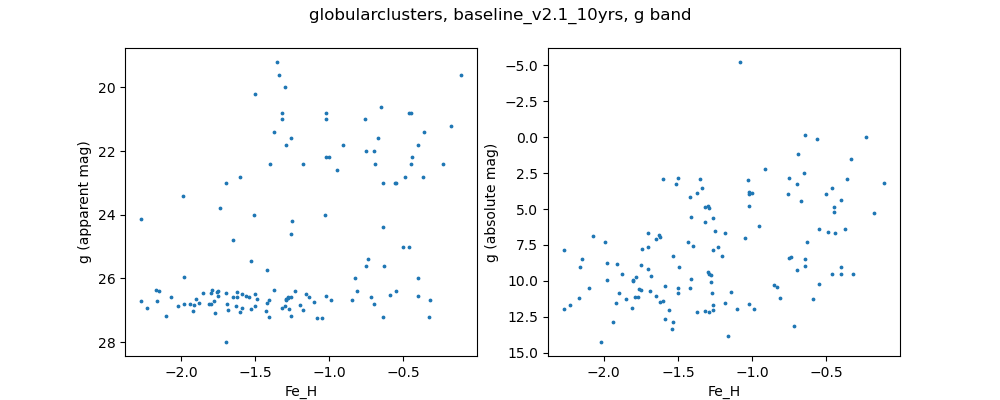} \\
    \includegraphics[width=0.49\textwidth]{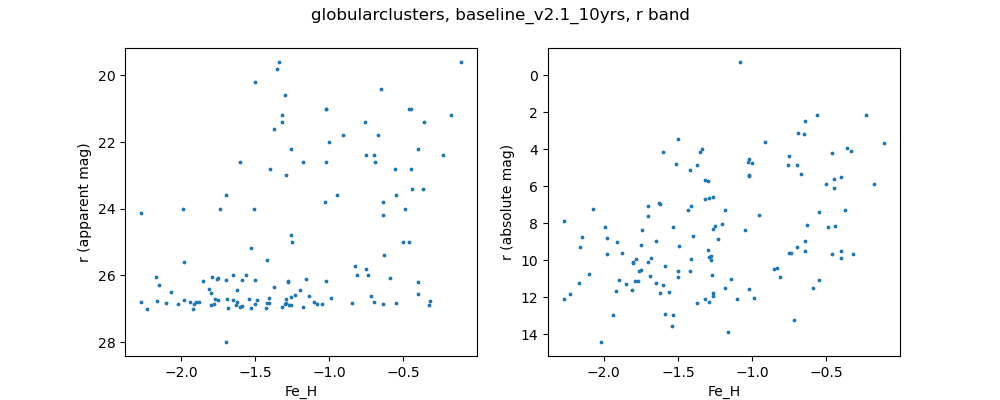}
    \includegraphics[width=0.49\textwidth]{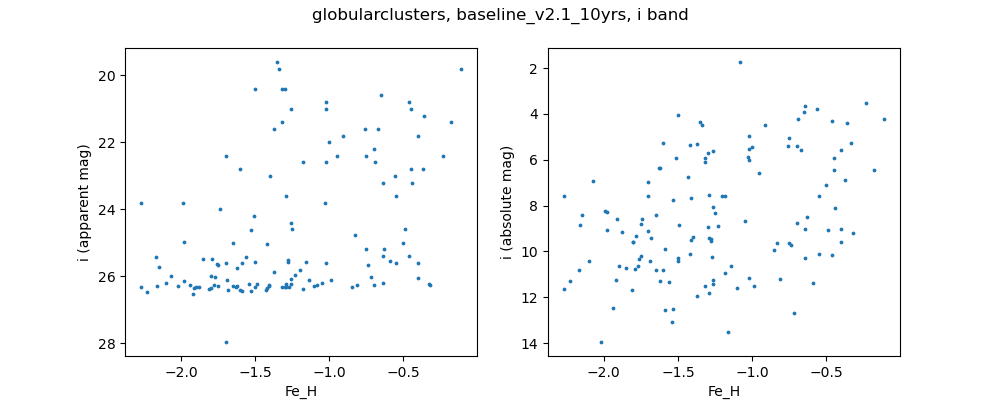} \\
    \includegraphics[width=0.49\textwidth]{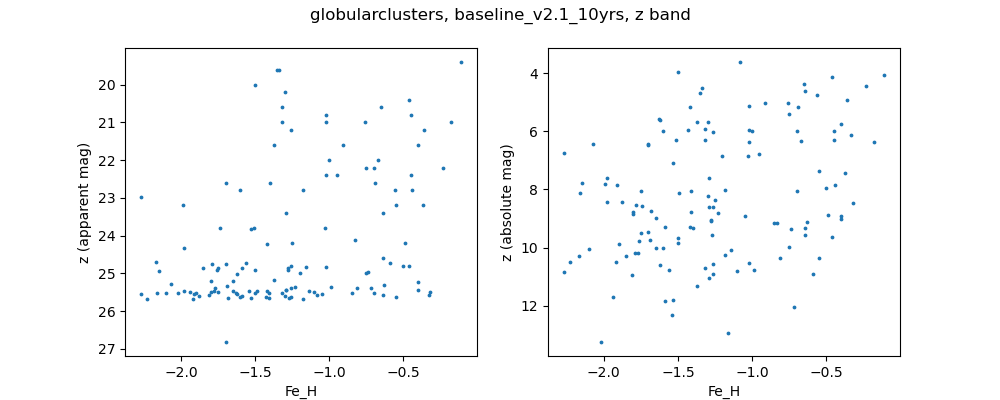}
    \includegraphics[width=0.49\textwidth]{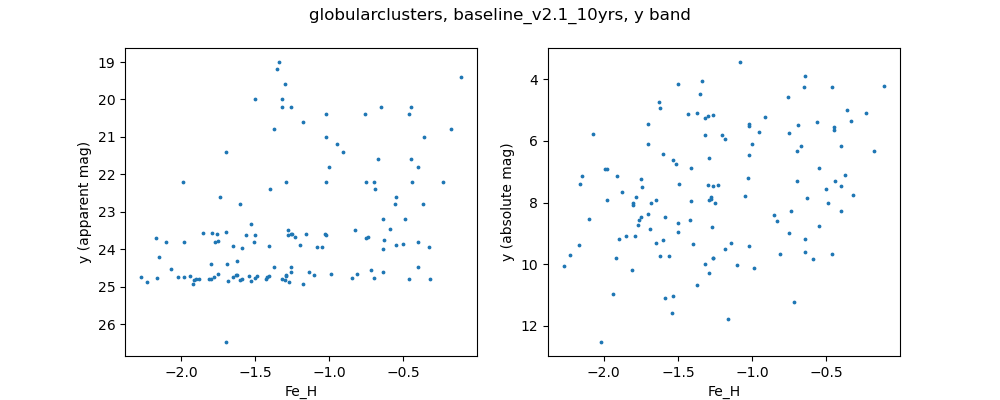}
    \caption{Maximum photometric depths to be reached in the outskirts of known MW GGCs, according to eqs.~\ref{eq:apparent} and \ref{eq:absolute} and assuming \texttt{baseline\_v2.1\_10yrs}. Results are presented for all LSST filters, in AB mags, as a function of [Fe/H].}
    \label{fig:MW_GGC_depth}
\end{figure*}

\begin{figure*}
    \centering
    \includegraphics[width=0.49\textwidth]{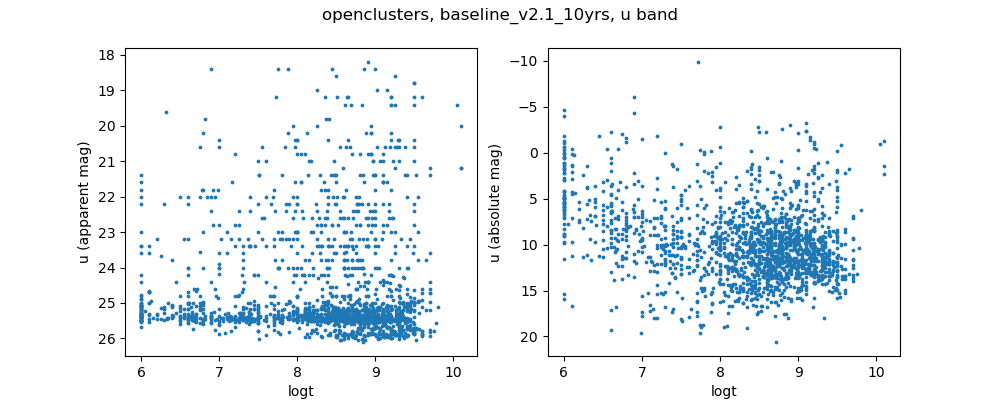}
    \includegraphics[width=0.49\textwidth]{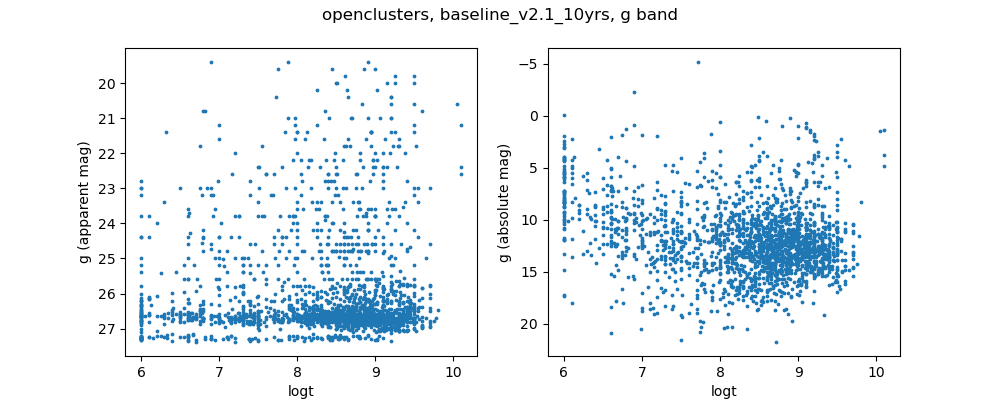} \\
    \includegraphics[width=0.49\textwidth]{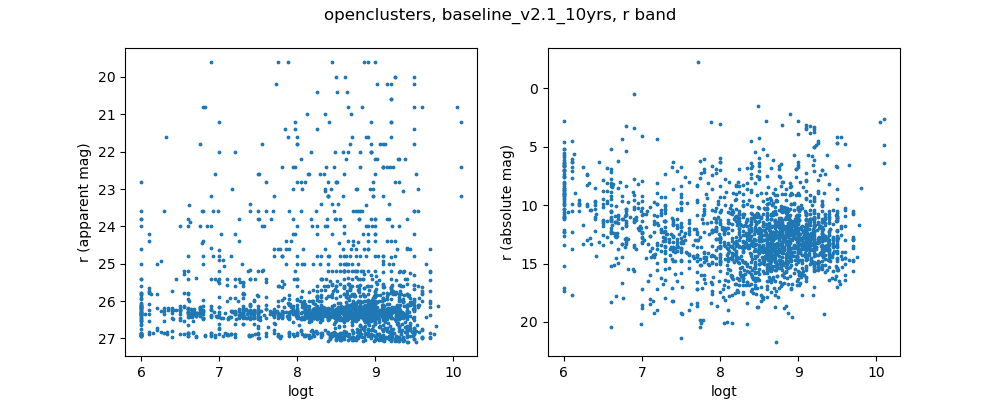}
    \includegraphics[width=0.49\textwidth]{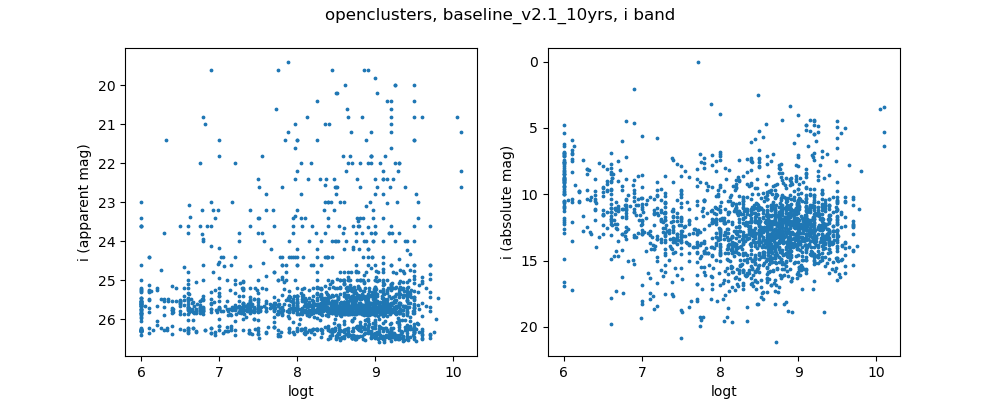} \\
    \includegraphics[width=0.49\textwidth]{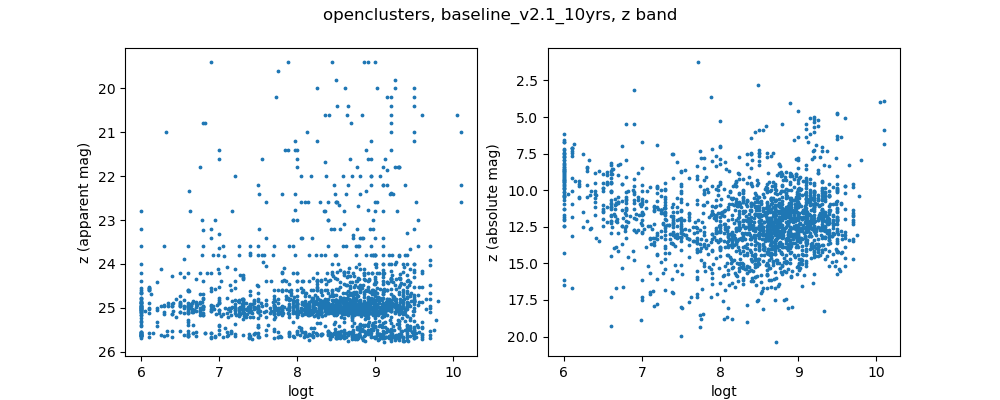}
    \includegraphics[width=0.49\textwidth]{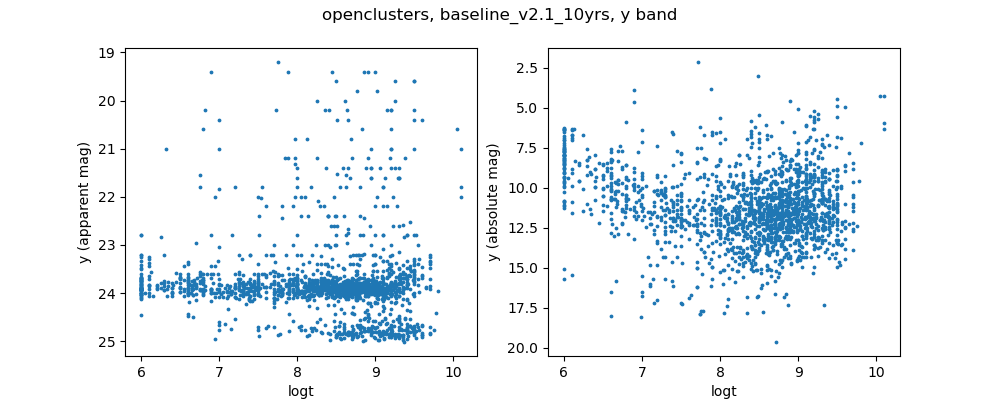} 
    \caption{The same as in Fig.~\ref{fig:MW_GGC_depth}, but for OCs as a function of log(age/yr).}
    \label{fig:MW_OC_depth}
\end{figure*}

All the $m_\lambda$ and $M_\lambda$ values are plotted in Figs.~\ref{fig:MW_GGC_depth} and \ref{fig:MW_OC_depth} against some of the main cluster properties listed in the original catalogs, namely [Fe/H] for GGCs and log(age[yr]) for OCs. Several aspects are evident in these plots: 
\begin{itemize}
    \item $m_\lambda$ values concentrate into two narrow strips close to the 25th-27th magnitude limits (depending on the filter). These two strips reflect the \texttt{CoaddM5} limit in uncrowded regions of the MW for the two possible numbers of visits (either $\sim810$ in the main WFD footprint, or $\sim250$ in the Galactic Plane mini-survey).
    \item There is an extended distribution of $m_\lambda$ to brighter values, as low as $\sim20$~mag. These points reflect the \texttt{CrowdingM5} limit in very crowded areas of the MW.
    \item Irrespective of their concentration/distribution in different $m_\lambda$ intervals, the clusters are well distributed in their intervals of [Fe/H] and log(age);
    \item The distributions in $M_\lambda$ appear quite promising, with significant numbers of clusters being sampled down to absolute magnitudes of $\gtrsim10$, for both OCs and GGCs. 
    \item Even more interesting is that significant numbers of OCs (of all ages) will be sampled to absolute magnitudes exceeding 15 (and even reaching 20 mag).
\end{itemize}
These latter points emphasize one of the greatest promises of LSST regarding star clusters: to sample faint features that have not been clearly documented in clusters of known age and metallicity, either by Gaia or by previous deep photometric surveys. We recall that the majority of the sample of field white dwarfs (WD) observed by Gaia starts at $M_G\gtrsim10$ and suddenly bifurcates at $M_G\gtrsim12$ \citep[corresponding to $M_u\gtrsim10$; see][]{babusiaux18} -- however Gaia did not observe WDs in clusters with $M_G\gtrsim13$. Moreover, we recall that the $15<M_i<20$ interval already corresponds to substellar objects, i.e.\ brown dwarfs. 


\subsection{Finding new MW clusters with LSST}
\label{sec:newMWclusters}

With unprecedented depth and homogeneous coverage, LSST will resolve  main sequence low-mass stars in clusters belonging to disk and halo of the MW, and stars at the top of the AGB belonging to the typical population of old and metal-poor star clusters out to $m-M \sim 28$ or 4 Mpc, assuming a range of magnitude equal to 1 magnitude below the brightest star in the AGB/RGB and a limiting magnitude equal to $g$=27. Despite the effect of crowding in the center of very distant objects, this will allow for a complete census of faint halo clusters and dwarf galaxies in the entire southern hemisphere in the large vicinity of the MW.
LSST will also provide astrometric measurements to four magnitudes fainter that provided by Gaia \citep[see section 3.7 of][]{lsstsciencebook}.
Methods for identifying new clusters and associations through the combination of photometry, parallaxes, and proper motions have recently been demonstrated by \citet{Kounkel:2020}, who also have extended the technique to very large samples with the application of machine learning \citep{McBride:2021,Prisinzano2022}. A few other methods have been refined using recent Gaia data releases, as nicely summarised in \citet{hunt21,hunt23}.

The resulting data will offer crucial insights into the interaction between the Galaxy and the nearest dwarf galaxies, such as the Magellanic Clouds, as well as the interactions among the dwarf galaxies themselves (see for instance \citealt{2015MNRAS.453.3568D}, \citealt{2016MNRAS.461.2212J}) and its group of satellites \citep{2020ApJ...893..121P}. This information will improve our understanding of the behavior of external star clusters along the interactions, and help determine the original host galaxy of the star cluster (see for instance \citealt{2019ApJ...875..154M} and references therein). Additionally, the data will provide clues about the evolution of galaxies and help to evaluate the bottom-up scenario for the formation of galaxies in the $\Lambda$-CDM model \citep{2005ApJ...635..931B}.

However, detecting star clusters in LSST presents a challenge due to the amount of data that will be produced. An estimation of about one billion stars down to $g$=27 is very realistic \citep[see, e.g.,][]{daltio22}, without accounting for galaxy/QSO contamination and regions very close to the Galactic plane, while using typical limits in color for expected stellar evolutionary phases in old and metal-poor systems (MS, RGB, AGB, and HB stars). Typically, in order to detect star clusters a matched filter is applied to the data sample (see for instance \citealt{2013ApJ...767..101B}, \citealt{2015ApJ...807...50B} and \citealt{2015ApJ...805..130K}), selecting stars belonging to a specific stellar population at a specific distance. Local peaks in the stellar population are then detected, and a list of candidates is generated, which is subsequently analyzed and duplicates removed, among other steps. In LSST, we expect to read positions, as well as magnitudes in the $g$ and $r$ filters and their respective photometric errors. {We note that the situation is somewhat different in current Gaia-based searches for new clusters \citep[e.g.][]{castroginard20,hunt23}, which very much rely on the measured proper motions -- although the problems related with the huge amount of data are essentially the same.

To handle this vast amount of data, the search for stellar clusters in the vicinity of the MW must leverage computational expertise and run searches in multiple regions and distances simultaneously to provide results within a reasonable time (in the worst case, a very few days).

Tests are currently underway with a pipeline based on detections of clusters of galaxies \citep{2021MNRAS.502.4435A}, which selects stars based on old and metal-poor isochronal masks at specific distances to detect both dwarf galaxies and faint halo clusters in the Local Volume. Codes have been developed to create simulations based on LSST's photometric system and depth, inserting stellar clusters in a MW foreground of stars. After simulating an area of the sky, the code detects the candidates, running many regions in parallel and concatenating the list of detections into a single catalog that includes parameters such as position, signal-to-noise ratio, size, and distance.

\subsection{Young Open Clusters, Moving Groups, and Associations}

Very young open clusters, which span the range of ages relevant to protoplanetary disk lifetimes ($< 10-20$~Myr; e.g., \citealp{Ribas2014}), allow us to probe a key stage in early stellar evolution. At these young ages, the dynamics of young, low-mass stars are governed by the interaction with their surrounding disks. This star–disk connection regulates the accretion of mass from the disk to the star \citep[e.g.,][]{Hartmann2016} and the shedding of angular momentum at the disk–magnetosphere interface \citep[e.g.,][]{Ireland2021}, and has a long-lasting impact on the fundamental properties of the final star. In addition, local changes in density and structure that are triggered by the accretion flow across the inner disk can reverberate on the migration pattern of close-in planets, impacting how these settle onto their final orbits \citep{Liu2017, Romanova2019}. 

The physical parameters of the youngest pre-main sequence (PMS) stars (e.g., spectral type, radius, magnetic field configuration, mass accretion rate $\dot{M}_{acc}$) are often difficult to measure precisely over a broad range of stellar masses and ages, due to a mix of spatially variable reddening and nebular  emission from H\textsc{ii} regions, observational uncertainties, and model systematics \citep[e.g.,][]{Hillenbrand2004}. As a result, detailed studies of star formation and early evolution have historically been restricted to relatively nearby open clusters and star forming regions \citep[$\lesssim$ 0.5--1~kpc; e.g.,][]{Luhman2012}, with complete membership censuses from the highest to the lowest masses (spectral classes $\sim$B to L) limited to the closest stellar associations and moving groups \citep[$\lesssim$ 0.1~kpc; e.g.,][]{Bell2017}. Over the last few years, significant progress has been achieved in cataloging all clustered young stellar populations within a distance of $\sim$1.5--2~kpc, thanks to the synergy between accurate Gaia astrometric data and unsupervised machine learning algorithms \citep[][and references therein]{Prisinzano2022}. However, comprehensive determinations of individual stellar properties are required in order to capitalize on these results and build a uniform picture of star formation history and kinematics across our Galaxy. Thanks to the combination of deep sky mapping, multi-wavelength characterization, and long-term observational baselines, Rubin LSST will allow us to address many of these issues with unprecedented accuracy.

For example, it is now possible to empirically measure the radii of stars through the combination of broadband spectral energy distributions, providing the stars' bolometric fluxes, together with parallaxes from Gaia and/or from Rubin itself, via the technique of ``pseudo-interferometry" \citep[see, e.g.,][]{Stassun:2016,Stassun:2017}. In addition, when spectroscopic surface gravity measurements are available, it is then also possible to measure the stellar masses as well \citep[e.g.,][]{Stassun:2018}. The ability to measure stellar radii and masses in this way opens the prospect of LSST as an unprecedented factory for the determination of fundamental physical properties of stars with known ages by the tens of thousands \citep{Stassun:2018}. 

\begin{figure}
    \centering
    \includegraphics[scale=0.9]{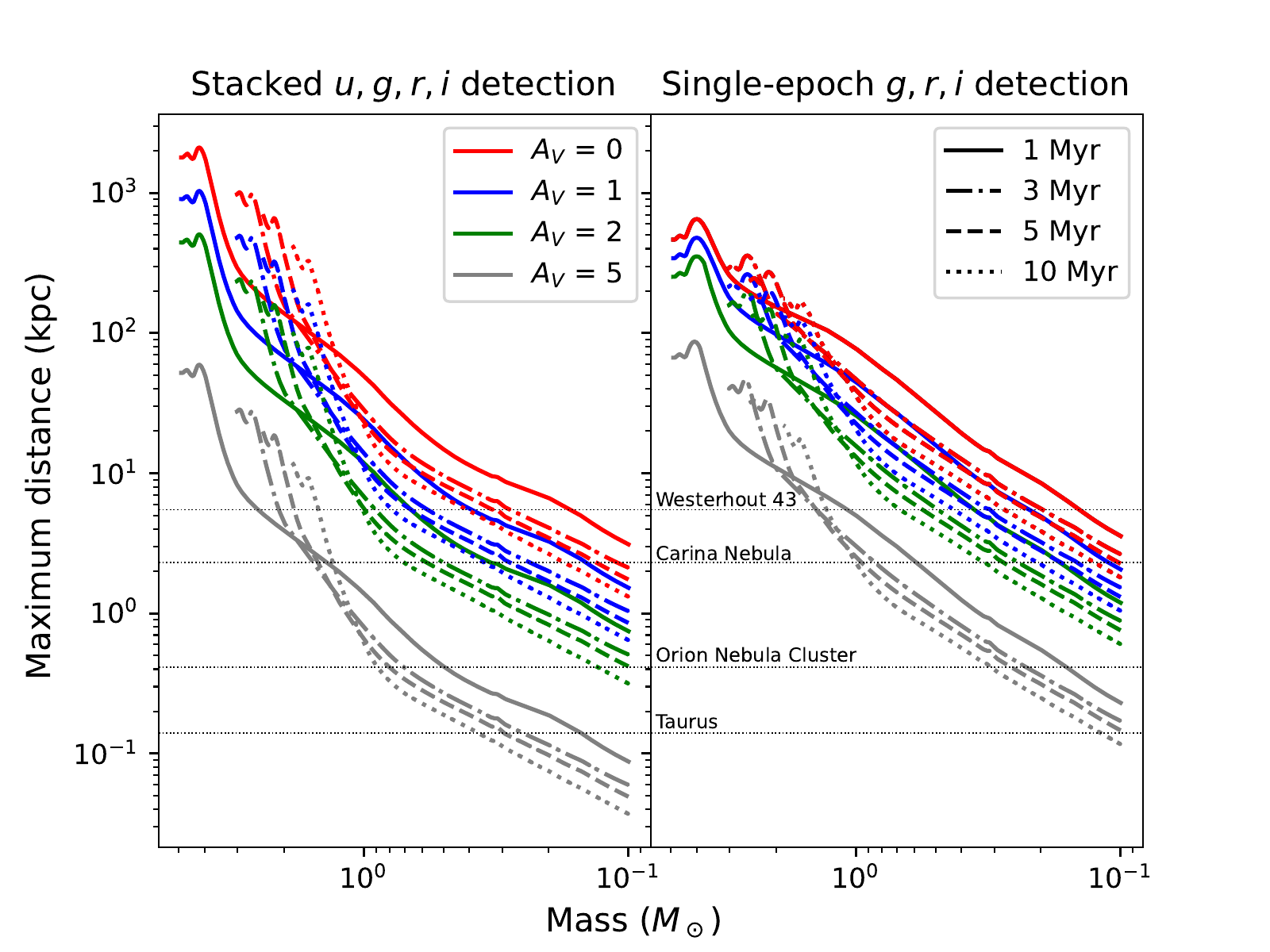}
    \caption{Projected maximum distance up to which young, PMS stars of different mass are expected to be detected in all of the $u,g,r,i$ filters at the end of the 10-year Rubin LSST survey (left panel), and in a single visit in $g,r,i$ filters (right panel), as a function of age (different line styles) and reddening (different colors; see legend). The distances to some well-known star-forming regions are marked on the diagrams as a reference for the scale. These projections were derived by assuming the single-image and coadded-image $5\sigma$ depths simulated for the baseline survey \citep{bianco22}, and converting them to expected limiting distances by using the MIST isochrones \citep{Dotter2016,Choi2016} with synthetic LSST photometry, and a sample range of ages and extinction values.}
    \label{fig:PMS_LSST_detection_limits}
\end{figure}

Studying the dynamics of disk accretion in young stellar objects is best done in the ultraviolet (UV) regime, which is sensitive to the energetic emission released by surface shocks that form at the location where magnetically-channeled material from the disk is transferred to the star at near free-fall velocities \citep{Calvet1998}. The photometric $U$-band (or $u$-band) provides an effective tracer of the excess emission driven by accretion shocks on top of the stellar photosphere \citep{Gullbring1998}, and when combined with optical photometry, it can be used to define accretion-sensitive color indices that can be converted to instantaneous measurements of the total accretion luminosity and $\dot{M}_{acc}$ \citep[e.g.,][]{Rigliaco2011,Venuti2014,Kalari2015}. The contemporaneous availability of multiple UV-optical filters is critical  for a proper estimation of stellar and accretion parameters \citep[e.g.,][]{Stassun1999,Venuti2021}, especially for disentangling the flux contribution due to accretion, the stellar activity component, and the effect of interstellar extinction \citep[e.g.,][]{vrba1993,Bouvier1993,Venuti2015}.

Key questions regarding the star-disk interaction stage of PMS evolution include for how long disk accretion persists, how the pattern of disk accretion is influenced by stellar properties such as mass and metallicity, and what connects routine, low-level accretion ($\dot{M}_{acc}\sim 10^{-8} M_\odot/yr$ around solar-type stars) to eruptive accretion events where $\dot{M}_{acc}$ can increase by 2--3 orders of magnitude \citep{Fischer2022}. Answers to these questions have significant implications for the angular momentum evolution of young stars, specifically setting the timescales of interest for interconnected mechanisms such as stellar winds and mass ejection \citep{Bouvier2014}. With an assumed Wide-Fast-Deep coverage of the Galactic plane and the many star-forming regions it encompasses \citep{Prisinzano2023}, Rubin LSST will have a transformative impact on these issues, by enabling simultaneous $u,g,r,i$ detection of young ($\sim$1--10~Myr) stars with relatively low extinction down to masses $\sim$0.1--0.2~$M_\odot$ (spectral types $\sim$M5--M4) to distances of up to a few kpc. This is illustrated in the left panel of Fig.~\ref{fig:PMS_LSST_detection_limits}, which represents the maximum distance that can be reached for a given limiting magnitude as a function of stellar age, stellar mass, and reddening, as presented in \citet{Damiani2018}. 

 A detailed analysis of the maximum distances that can be reached for a 10 Myr old star of 0.3~M$_\odot$, by adopting four different Rubin Opsims, including the case with the Wide-Fast-Deep cadence extended to the Galactic Plane, is reported in 
 \citet[][see their Fig. 5]{Prisinzano2023}. Rubin LSST simulations adopting their metric show that, with the deepest cadences, distances up  to  9-14 kpc can be achieved. Such limits depend on the stellar ages, masses and reddening but are also strongly affected by the crowding effects. 

The extent and depth of this sky mapping will allow uniform, quantitative measurements of the prevalence and strength of accretion activity for PMS stars of varying mass and age in different environments than the Solar neighborhood. When limiting the selection to the optical $g,r,i$ filters, young, late-type M-stars are expected to be detected down to distances of a few kpc in a single visit (Fig.~\ref{fig:PMS_LSST_detection_limits}, right), and up to a factor 3--4 deeper in the final coadded stack at the end of the survey. 
For intermediate-distance regions, this combination of depth and multi-wavelength information will enable cluster memberships to be refined down to the lowest masses; this is essential for accurate reconstructions of the stellar initial mass function (IMF) and to test its universality \citep{Offner2014}. At distances $>$1--2~kpc, the homogeneous detection of M-type stars will provide an invaluable tool to statistically identify and characterize new clustered PMS populations, by taking advantage of the peculiar color properties of M-dwarfs in $g,r,i$ filters \citep{Damiani2018,prisinzano2018,Venuti2019} and of the brightness contrast between young, M-type PMS stars and more evolved M-type field dwarfs \citep[e.g.,][]{Siess2000}.

The peculiar color properties of M-dwarfs, which represent more than 80\% of stellar cluster members \citep{lada06}, also allow us to estimate individual reddening values. These are crucial to derive effective temperatures and stellar luminosities and then stellar ages by comparing them with stellar isochrones in the HR diagram. 
In turn, this enables analyzing the age distributions of stars in young clusters, which is necessary to address the still-unresolved question of the age spread and to impose strong constraints on the physical mechanisms and timescales involved in the process of cluster formation \citep[e.g.,][]{tassis04,ballesteros07,palla05}.

While typical changes in the young star--inner disk environment are triggered on comparable or shorter timescales (days to weeks) than the expected observing cadence for LSST fields \citep[e.g.,][]{Costigan2014, Flaischlen2022}, the long observational baseline over which young open clusters and star-forming regions will be regularly sampled will be critical to assess the interplay between inner disk and outer disk dynamics, and to investigate the causal link between small-scale disk processes and large-scale eruptive behaviors that can develop over months to years \citep{Bonito2023}. 

The homogeneous, multi-wavelength database that will be assembled over the entire Rubin LSST lifetime will also enable unprecedented evaluations of stellar magnetic cycles during the PMS phase, uniform measurements of stellar rotation rates, the identification of new PMS eclipsing binary systems in clusters (invaluable benchmarks for validating PMS evolutionary models; e.g., \citealp{Gillen2020,Stassun:2022}), and potential detections of transiting substellar objects \citep[e.g.,][]{Stassun:2006}.
With regards to stellar rotation and the evolution of stellar angular momentum, recent work demonstrates the power of combining time-series photometry of large samples of young stars whose ages are known by virtue of their membership in clusters and associations \citep[][see Figure~\ref{fig:angmom}]{Kounkel:2022}; LSST will dramatically expand the samples and more completely fill in the parameter space of stellar age, masses, and metallicity. 

\begin{figure}[!ht]
    \centering
    \includegraphics[width=\linewidth,trim=30 20 150 132,clip]{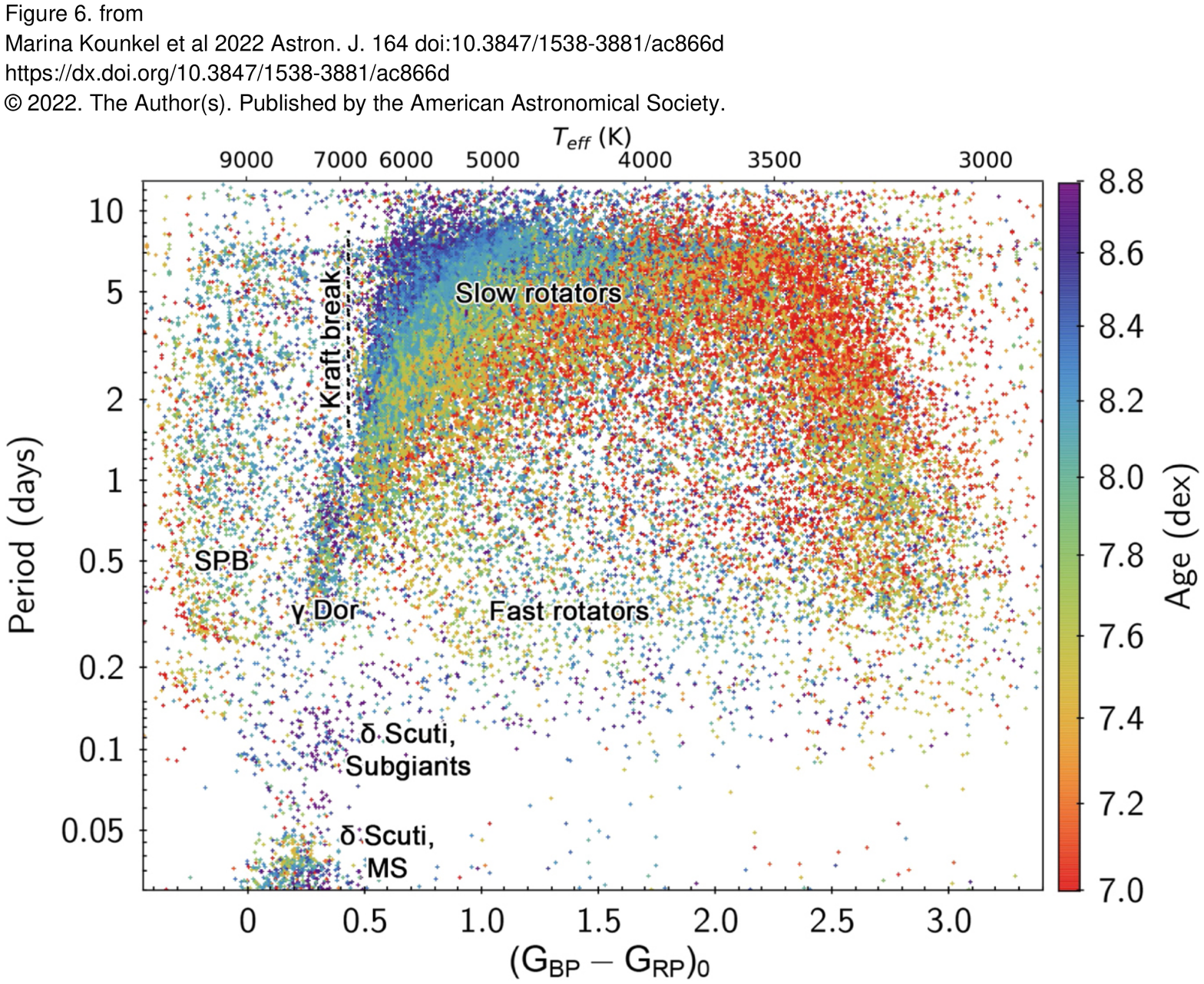}
    \caption{Recovered distribution of periods of the stars in the sample as a function of Gaia color, extinction-corrected, color-coded by their age, annotated with the features of the parameter space. Sources redder than the labeled ``Kraft break" have the dominant mode of variability being caused by rotation. They can be separated into slow rotators (sources that show a clear evolution of their periods with age), and fast rotators, having typical periods $<$1 day. Sources bluer than the Kraft break are commonly pulsators. $\gamma$~Dor sources have pulsation frequency comparable to the stellar rotation period, consisting only of the main-sequence stars. $\delta$~Scuti stars in the age-limited sample form a bimodal distribution, with the main-sequence $\delta$~Scuti variables having periods shorter than 1.2 hr, and the subgiants having periods longer than 2 hr. Slowly pulsating B-type variables (SPB) are found among the early-type stars. Figure from \citet{Kounkel:2022}, reproduced by permission of the AAS.}
    \label{fig:angmom}
\end{figure}

\subsection{Star clusters as stellar evolution tools}


Star clusters in the Milky Way that are fully resolved into stars  are invaluable tools for calibrating stellar evolution models. The evolved stars within the clusters are expected to closely track theoretical isochrones in proportions that directly follow from the amount of time spent in different evolutionary phases, with only minimal dependence on the initial mass function (IMF). Particularly important in this regard are the intermediate-age star clusters, i.e. those old enough to develop an extended RGB sequence, but not old enough to be classified as ancient GGCs. Their main sequence turn-offs in the HR diagram display the characteristic hook caused by convective cores. They are also of primary importance for stellar population models, because they probe most of the history of our Galaxy, over the complete age range from $\sim1$ to 8 Gyr.  

The stars on the Main Sequence of such clusters have small convective cores, which brings a significant amount of uncertainty in the age-dating of the clusters. Indeed, stellar models with different amounts of convective core overshooting may change the inferred ages by factors exceeding 50\%. 
In addition, in recent years it has been recognized that the presence of fast rotators has a significant impact both in the definition of the ``typical amount of convective overshooting'' experienced by stars with all masses larger than $\sim1.2$\Msun\ \citep{costa19}, and in the stars' estimated ages. Indeed, rotation increases core masses and luminosities at the end of the main sequence in a way that partially mimics overshooting \citep[see e.g.][]{ekstrom12}. A significant spread in rotational velocities causes main sequences with broad turn-offs at intermediate ages and split sequences in the youngest clusters \citep[e.g.][]{kamann20,kamann23}, but leaves no trace in the red giants, except for their increased core sizes and changes in the surface abundances of CNO elements. Their signature in color-color plots is tiny, although possibly detectable in star clusters with a very uniform reddening, and with a high fraction of fast rotators \citep{girardi19}.

Most of the present inferences regarding rotation derive from star clusters in the Magellanic Clouds (MCs), thanks to very precise HST photometry and large stellar populations in these objects. However, the study of rotation in the MCs is challenging given the difficulties in obtaining accurate spectroscopy in crowded clusters at $>50$~kpc distances. Fortunately, the phenomenon has also been identified in a few very nearby clusters \citep{brandt15, gossage18,cordoni18}, and their characterisation in other MW clusters might improve further with better membership information and dedicated spectroscopic studies.
It is obvious that LSST will help to significantly expand the useful database of precise photometry and good memberships for these MW star clusters. In addition, the use of 6-band photometry is expected to produce much better extinction maps, hopefully further clarifying the features caused by fast rotators in the CMD.

Another controversial subject in recent years is the initial-to-final mass relation (IFMR) of stars in the 1 to 8 \Msun\ interval. While this relation has been regarded as a close-to-linear relation with little dependence on metallicity \citep[e.g.][]{weidemann00}, claims have emerged of a significant non-monotonic behaviour, or a kink, in the IFMR at near-solar metallicities \citep{cummings18, marigo20}. This may have significant implications for understanding the evolution of asymptotic giant branch (AGB) stars, which provide a fair fraction of the newly-synthesized metals in galaxies, and of their near-infrared luminosities. Progress in this sense requires (1) a significant expansion in the detection and mass measurements of WDs in star clusters, and (2) eventually checking the mechanism that causes the kink using AGB stars in clusters \citep{marigo22}. Candidate AGB stars in MW clusters are presently very rare, and lack some relevant information such as their chemical types (roughly corresponding to the M/S/C spectral classification) and long-period variability. These are two areas in which the precise 6-band time-series photometry of LSST, added to existing near-infrared photometry, will certainly produce some improvement.

Also important in these star clusters is the presence of products of close binary evolution, in particular the blue stragglers and hot subdwarfs, which form in a few subtypes and via (at least) a couple of channels whose relative importance are still not clear \citep[see e.g.][]{heber09,leiner21,jadhav21}. Such stars are rare but the reliability of present catalogs has very much improved thanks to Gaia \citep{rain21,leiner21,jadhav21,geier19,culpan22}. These types of stars are more easily detected in blue/UV filters, but the Galactic disk extinction is probably hiding a significant number of them from our view. The LSST deep $u$-band photometry in the near-disk region will be invaluable to increasing the size of the statistical sample in the MW.

Finally, double-lined eclipsing binaries (DLEB) allow the precise determination of stellar masses and radii of their components, including for stars with known ages and representing alternate stellar evolution pathways such as red stragglers and others \citep[see, e.g.,][]{Stassun:2023}. The few DLEBs presently known as cluster members indeed set strong constraints on the global properties of the host cluster, such as age and distance, and eventually on their initial helium content \citep[e.g.][]{thompson10,meibom09,brogaard21}. LSST may significantly increase the numbers of known eclipsing binaries in clusters by factors of 1.7 to 3 \citep[for globular clusters and open clusters, respectively;][]{geller21}, eventually leading to significant progress in this front.

\subsection{Galactic Globular Clusters}

Ninety percent of the known Milky Way globular clusters are located within the LSST footprint.
As can be seen in Figure~\ref{fig:MW_GGC_depth}, LSST photometry will reach well below the main sequence turnoff at least in the outskirts of almost all of these GCs, while also reaching the hydrogen burning limit in a few clusters.
So far, only Gaia provides a larger homogeneous sample of resolved star photometry in GCs, however  LSST will reach five magnitudes deeper, and in six filters, as opposed to the 3 Gaia filters. 

We have been able to learn a lot based on the photometric depth of HST's dedicated surveys on Galactic GCs \citep[e.g.][]{acssurvey,hugs}. However, the coverage of the HST observations did not allow for wide field studies, encompassing the clusters as a whole. In contrast, the nearly all sky-coverage of LSST will provide aspatially comprehensive analysis of the Milky GC system, allowing us to study GCs from their central parts (dependent on crowding, etc.) to their outskirts. 
The largest homogeneous sample of wide-field photometry for GCs is from \citet{2019MNRAS.485.3042S} which covers 48 GCs and makes use of a wide variety of instrumentation. In contrast, LSST will be homogeneous and deeper. For inner halo GCs, Gaia has allowed detailed explorations of GC outskirts \citep[e.g.,][]{2019MNRAS.485.4906D,2020MNRAS.495.2222S,2021MNRAS.507.1127K}. Indeed, as illustrated and quantified by \citet{cantatgaudin23}, the Gaia catalogues become severely incomplete in the central parts of globular clusters.

The deep $u$-band photometry provided by LSST will allow the identification of N-rich stars and other features that can reveal multiple stellar populations in GCs. \citep[e.g.][]{2011A&A...525A.114L}.
This can in turn be used to probe large scale radial variations and abundances in the different stellar populations of Galactic GCs.
This is of foremost importance, as virtually all globular clusters studied in depth show evidence for multiple populations \citep[see review by][]{2018ARA&A..56...83B}. It is worthwhile to note that the community has not yet reached a consensus as to what is the key driver for the appearance of  multiple stellar populations in clusters.
Moreover, it will be possible to look for N-rich giants outside of GCs and trace back field populations to previous generations of disrupted clusters (\citealt{2011A&A...534A.136M,schiavon17}).




Another important contribution will be on analysing the outskirts of clusters and examining the properties of their extra-tidal stars and searching for evidence of tidal streams. LSST will enable the detection of even larger samples of stars than has been possible in the Gaia era \citep{2019NatAs...3..667I,2019ApJ...887L..12B}.

LSST will also enable a detailed study of variable phenomena including: Pulsating variable stars in clusters, RR Lyrae stars in globular clusters, and Cepheid variables in open clusters, i.e., in populations of known ages and metallicities (see \citealt{2022arXiv220804499H} for a further discussion).

\subsubsection{Clusters in The Galactic Bulge and Plane}

The Galactic bulge and plane host roughly 2/3 of the Galaxy's globular clusters, yet are burdened by the perennial problems of high extinction and extreme crowding.  However, the cluster population is innately quite interesting; this region hosts some of the most metal poor and metal rich of the Galaxy's clusters, as well as two massive clusters that are unlike any others known, hosting complex populations exhibiting multiple peaks in [Fe/H] as well as evidence of distinct bursts of star formation separated by many Gyr.  Terzan 5 \citep{ferraro09,origlia11} has 3 metallicity peaks ranging from $-$1 to +0.5 dex; Liller 1 \citep{ferraro21} has multiple age bursts including 12  and 1 Gyr. 

The Blanco DECam Bulge Survey \citep{2020MNRAS.499.2340R,2020MNRAS.499.2357J} has shown how multicolor photometry, combined with Gaia astrometry along with  application of the vector point plot to clean the globular cluster color-magnitude diagrams yields dramatically improved results.  LSST will cover all of the bulge clusters with multicolor photometry, dramatically better image quality, and Gaia matching.  This trifecta of multicolor imaging, astrometry, and radial star counts to confirm cluster membership has already paid off.
While new imaging technology such as the Variables in the Via Lactea (VVV) survey have uncovered many new candidate globular clusters in the bulge and plane, applying multicolor photometry and astrometry has shown some of these clusters are not real \citep{2020MNRAS.499.2340R}.  Combining $u$-band photometry from the Blanco DECam Bulge Survey with Gaia astrometry has led to clear detection of multiple populations in bulge globular clusters \citep{kader22} again emphasizing the value of deep $u$-band observations and multicolor imaging that LSST will offer.

The combination of outstanding multiband photometry with Gaia astrometry, as well as imaging in the infrared, will produce data of remarkable quality.  Proper motion vector point diagrams can be used to clearly separate cluster stars from field members; additional Gaia data releases will only lead to dramatically improved results.

LSST will have some significant advantages in the bulge and plane.  The uniformity of photometry, potential development of tools that provide high resolution dereddening of the data by taking advantage of the multicolor photometry will be of great use in studying the clusters and their field populations.   The multicolor data will yield robust metallicity and (for those clusters with sufficiently deep photometry) age estimates.  If imaging can span multiple epochs, new variable star populations will be uncovered as well.

In order to exploit fully the promise of LSST, the astrometric solution of Gaia will need to be extended to faint magnitudes, so that eventually photometric cleaning may be applied to stars fainter than the Gaia limit of $g\sim21$.  If excellent astrometry is achieved to $g=24$ or fainter, we gain leverage on many other important problems, such as the search for ultra-faint dwarf galaxies 

\subsubsection{Multiple Stellar Populations}

Over the past two decades the presence of Multiple Stellar Populations (MSPs) in Galactic Globular Clusters has been well established as the norm (see \citealt{2018ARA&A..56...83B} and \citealt{Gratton2019} for recent reviews on the topic). In essentially all well studied GCs at least two populations have been identified: one that has similar abundance patterns to surrounding halo stars and a second population that tends to be enriched in Helium and, more importantly, presents negative correlations between some light elements (e.g.: C, N, Na, O). Spectroscopic studies have provided an in-depth picture of these chemical differences; however, such studies are limited to a small fraction of GC stars \citep{Carretta2009}. Meanwhile, many groups have been successful in using space- and ground-based photometric observations to separate and characterise the populations in a large number of clusters \citep{Piotto2015,Lardo2011,Hartmann2022, 2023MNRAS.tmp..269L}. 
However, so far wide-field studies of MSPs only encompass a limited number of clusters.

The LSST project has a great potential to expand this field. The majority of spectral features that can be used to differentiate between populations in GCs are concentrated in the bluer parts of the spectrum and the SDSS u-band is a valuable tool for distinguishing them. Because the majority of the Galactic GCs will be inside the footprint of LSST, a homogeneous analysis of a large set of clusters will be possible. The expected depth of the multi-epoch survey (well bellow the turn-off in most clusters) will also allow for a view of MSPs from the main sequence to the red giant and asymptotic giant branches. The wide field nature of the survey will be important in understanding how the populations behave far away from the cluster center and synergies with Gaia can help expand our understanding of the dynamics of the different stellar populations (see e.g. \citealt{martens23}.) 









\section{Semi-resolved Star Clusters in the Local Group}


By `semi-resolved star clusters', we mean objects similar to the star clusters in the Magellanic Clouds (MCs).  Such clusters can be fully resolved into stars only through diffraction-limited imaging with space telescopes like HST, or adaptive-optics-assisted imaging from the ground.  On the other hand, imaging with $\sim$1-arcsec resolution from the ground already allows detailed studies of the brightest stars and the variables, as well as integrated photometry.  Therefore, such clusters lie in a transition region between ``resolved stars'' and ``integrated light only''. The advantage of LSST in regard to these clusters is that, for the first time, the same survey (and same filters) will cover huge numbers of clusters over the entire resolution scale going from resolved MW clusters to unresolved star clusters beyond the Local Group. This study will be even more powerful when combined with Euclid \citep{2022zndo...5836022G}.

With the expected resolution of Rubin (median seeing of $\sim 1$ arcsec, best seeing of $\sim 0.7$ arcsec), we expect to measure the photometry of the brighter stars in semi-resolved clusters, especially in the cluster outskirts.

Single epoch observations with LSST will be deeper ($r = 24.5$) than most existing surveys of the MCs (STEP, SMASH) and will reach the main sequence turnoff at old ages if the effects of crowding are ignored.
Given the large number of epochs with LSST, the best images will have significantly better seeing ($\sim 0.7$ arcsec) than these surveys ($\sim 1.1$ arcsec).

We can better appreciate the likely role of Rubin by examining the observational situation of the very populous clusters in the Magellanic Clouds (MCs). A hundred of them have HST photometry in at least 2 filters \citep[e.g.][]{milone23}, and can be regarded as references for the definition of an age and metallicity sequence for MC stellar populations. Since the HST resolution is unbeatable (but for JWST), these clusters can be used to test the LSST photometry pipeline, as well as the analysis
tools, results and data products.   
For clusters without HST imaging, the targeted adaptive optics imaging survey VISCACHA \citep{2019MNRAS.484.5702M,2020IAUS..351...89D} provides a first idea of the improvements that will be possible with LSST. VISCACHA reaches a similar depth to a single LSST epoch but with better image quality ($\sim 0.5$ arcsec), with available data for more than 230 clusters in the outskirts of LMC, SMC and Magellanic Bridge. On the other hand, Rubin will provide imaging in six bands, which is better for constraining metallicity and reddening. 

Moreover, LSST will significantly improve the time-series data, providing the most extensive survey of variable stars in the MCs since OGLE IV \citep{udalski15}. 

\subsection{Census and properties of star clusters in the Magellanic Clouds}

The star cluster system of the MCs has always been one of the most studied for the following reasons: (i) they are sufficiently close to allow us to resolve their individual stars from the ground; (ii) a significant number of star clusters are sufficiently massive to populate the evolutionary phases following the Turn-Off; (iii) the SCs in the Magellanic system are present in great numbers. Indeed, the updated catalogs by \citet{Bica2008} and \citet{Bica2020} include more than 4000 objects, comprising star clusters and young associations.
A significant sub-sample of MC star clusters is also as populous as the old GCs of the MW, but with the difference that they span a wide range of ages, from a few Myr to 10 Gyr, making them suitable to probe a broad parameter space and to study how their physical and structural properties evolve during their lifetime. Furthermore, the Magellanic Clouds have massive young and intermediate age clusters that are not seen in the Milky Way.
Additionally, since we observe the Clouds from outside of the galaxies and due to their lower metallicity, the extinction effects are significantly less than for much of the Milky Way.

In this context, the LSST survey will provide us with unprecedented data with respect to what is currently present in the literature. The major advantage will be the availability of deep, precise and homogeneous photometry in six bands. This unique data set will allow us to achieve the following goals:

\begin{itemize}
\item{\bf Completion of star clusters census in the MCs system.} It has been suggested \citep[see e.g.][and references therein]{Gatto2020} that a conspicuous number of star clusters is still missing, especially in the LMC. This is due to several factors: i) the too-shallow photometry used to search for star clusters in certain regions of the MCs, which hampered the ability to detect low-mass faint star clusters and biased the detection towards young and intermediate-age star clusters compared with the old ones of similar mass (ages larger than $\sim$5-7 Gyrs) which are fainter; ii) crowding in the bar of the LMC and the main body of the SMC, resulting again in biasing the detection towards more massive star clusters; iii) inhomogeneous photometry in different regions. 

It is difficult to estimate the number of new star clusters which will be detected based on the LSST deep photometry, however, Fig. 19 by \citet{Gatto2020} seems to suggest that several hundred star clusters are still missing in the LMC only, especially in the Bar and at galactocentric radii $>$ 3 deg (a point where the coverage of the available wide-field surveys starts to be incomplete and more in-homogeneous).   

\item {\bf Measurement of the star cluster parameters (age, distance, metallicity, reddening).}

The deep LSST photometry will allow us to reach some magnitudes below the turn-off magnitude of the oldest MC star clusters. This means that we can exploit their CMDs to accurately estimate the age, distance, metallicity and reddening of almost all the star clusters in the MCs, after a proper decontamination of field stars. It will also be possible to investigate the luminosity and mass functions of the clusters down to very low stellar masses. For the most crowded clusters, we will take advantage of the single images obtained during the best-seeing conditions, which will be deep enough to reach at least two magnitudes below the turn-off of the oldest cluster. The use of six bands in the procedure will allow us to some extent to remove the degeneracy between age, reddening and metallicity. 
The integrated photometry of MC star clusters can be directly compared to those of star clusters in more distant galaxies.

\item {\bf Determination of star cluster structural parameters:}

The LSST data will allow us to build accurate surface brightness profiles (SBPs) which contain information about the internal structure of the clusters. In particular, for each cluster, by fitting the SBP with Elson or King profiles \citep[][]{Elson1989,Elson1991,King1962} it will be possible to measure the core (and tidal) radius, the peak SB and the total luminosity. In turn, these data can be used to estimate the mass of the cluster by means of, e.g., Monte Carlo experiments on synthetic CMDs.

\end{itemize}

In turn, the measurements described above will allow us to explore the following issues, some of which represent long-standing open problems:

\begin{itemize}
    
\item {\bf Age gap puzzle of the star clusters in the LMC.}

The existence of the so-called “age-gap”, an interval ranging from 4 to 10 Gyr in the
LMC, which was thought to be almost totally devoid of star clusters \citep[e.g.][]{DaCosta1991} has puzzled researchers for decades now. However, recent investigations based on relatively deep photometry have shown that several faint candidate star clusters in regions of the LMC not investigated before can have ages falling in the age-gap \citep{Gatto2020}. In addition, a re-analysis of the CMDs of the star clusters KMHK 1592 and KMHK 1762 based on deep photometry showed that both star clusters fall inside the age-gap, at odds with previous age estimates based on too-shallow photometry \citep[][]{Piatti2022,Gatto2022_KMHK1762}. These findings seem to suggest that the age-gap is not a real physical feature, but the result of an observational bias, originated by the combination of too-shallow photometry carried out in the literature, the lack of a proper decontamination of field stars and an incomplete investigation of the LMC outskirts, where the lower stellar density of the field permits the detection of even faint and sparse star clusters.

In this context, the contribution of the LSST survey will be to resolve any remaining doubt about the age-gap explanation, as it will allow obtaining both a complete census of the star clusters in the LMC outskirts, with the possible detection of old and faint star clusters not detected before, and to obtain CMDs deep enough to revise the ages of thousands of star clusters whose ages can in principle be very inaccurate, due to the shallowness of the photometry. 

\item {\bf Star Formation History of star cluster systems and comparison with the field.}

The completeness of the star cluster census and the accuracy of age and metallicity estimates that will be possible with the LSST data will allow us to put significant constraints on the star cluster formation history and to compare it with the field star formation history. In this way, it will be possible to test one of the paradigms of star formation, which forecasts that a significant  fraction of stars is formed within star clusters \citep{Lada&Lada2003}.   
At the same time, the comparison between the star cluster formation histories of the LMC and SMC could allow us to verify whether the star cluster formation history has been triggered by multiple close encounters between the two MCs (and possibly between the MCs and MW), as recently suggested for the field stars \citep[][]{Massana2022}, as well as to verify epochs of variation in the age-metallicity pattern followed by increases in the star formation rate, in order to understand the complex history of the system.

\item{\bf Star clusters 3D geometry, radial gradients.}

The individual homogeneous estimates of metallicity, distance and reddening for the whole sample of star clusters in the MCs will allow studying of the tridimensional geometric distribution of the star cluster systems and compare it with that obtained from other tracers, such as the RC stars, as well as Cepheids and RR Lyrae variables.   
It will be possible to investigate possible gradients of age or metallicity
 coupled with radial migration or the interaction history of the MCs \citep[e.g.][]{Dias2016}.
 Moreover, for clusters with precise radial velocity measurements, the proper motion given by the LSST multi-epoch data will provide a complete 6D space-phase vector (coordinates, distance, proper motions and radial velocities). This can lead to strong conclusions, regarding expanding or stretching structures in the MCs outskirts \citep[which can be compared to dynamical models; e.g.][]{2022MNRAS.512.4334D}, as well as possibly clusters with different velocity distributions than of the host galaxy.

\item {\bf Physical processes governing the MC star clusters internal structure.}

It will be possible to explore the evolution of the ensemble SC's structural parameters mass and radii with age and position in the galaxies. In particular, it will be possible to investigate the dynamical status of each star cluster and e.g. individuate the physical mechanisms that induce the observed increase of the core radius after 0.3--1.0 Gyr \citep[e.g.][]{Mackey2003,Gatto2021}. The data available for thousands of star clusters will allow us to study the mass-size relationship almost over the entire range of encompassed by the star clusters in the MCs, from hundreds up to hundreds of thousands of solar masses.

\end{itemize}

\subsection{MC clusters contribution to studies of stellar evolution:}

MC clusters have been very important in the stellar evolution theory because of the presence of very populous clusters of young and intermediate ages, unlike the old classical globular clusters of the MW and the generally low mass Galactic population of young clusters. Some of the MC clusters contain impressive numbers of evolved stars, such as for instance NGC~1866 with its 24 Cepheids \citep{welsh93,musella06}, and NGC~419 with its $\sim20$ carbon stars \citep{frogel90}. Even the most populous clusters in the MW and in M31 do not host more than one example of such objects \citep[see e.g.,][]{riess22,marigo22,senchyna15,girardi20}. The most populous MC clusters have been target of HST proposals, revealing surprising features such as broad MSTOs \citep{mackey08,milone09}, split main sequences \citep{milone16,correnti17}, dual red clumps \citep{girardi09} and UV-dim MSTO stars \citep{milone23}, nowadays attributed mainly to the presence of a large fraction of fast rotators among the early-type stars of these clusters \citep[e.g.][]{2020MNRAS.492.2177K, 2023MNRAS.518.1505K, martocchia23}. Although LSST photometry cannot compete with HST photometry for the densest clusters, it will cover the entire LMC and SMC areas with extreme homogeneity, multiple filters and long time series, hence making it possible to:
\begin{itemize}
    \item Improve the correction of their foreground extinction, and the contamination from the LMC and SMC fields. Indeed, this is a weak point of HST photometry, for which the observations are often available in only 2 filters and covering too small areas to allow proper decontamination of the cluster photometry from the LMC/SMC field. Narrow CMD features and accurate star counts are an essential ingredient for stellar evolution studies using star clusters. 
    \item Look for the signatures of fast rotation (broad turn-offs, split main sequences, UV-dim stars) in sizeable samples of clusters including less populous objects, then allowing to explore open questions like: Are fast rotators more frequent in the most massive clusters than in small clusters and in the general galaxy field? Are slow and fast rotators truly coeval, and hence a single stellar population \citep[e.g.,][]{correnti21}? Does the stellar rotation occur with a preferential orientation inherited from the molecular cloud from which the star cluster was formed \citep[e.g.,][]{kamann19}? What is the role of close binary evolution \citep[e.g.,][]{wang20,kamann21}?
    \item Increase the numbers of variables in MC clusters, possibly adding constraints to the evolution of convective cores. For instance, fast rotation gives origin to larger H-exhausted cores at the end of the main sequence, hence causing a spread in the mass-luminosity relation of evolved stars. This spread could be documented in a number of ways, in particular 1) using the mass-radius relation derived from double-line eclipsing binaries, whenever one of the components is already evolved away from the zero-age main sequence; and 2) using mass estimates from Cepheids \citep{marconi13,costa19}. 
\end{itemize}

\subsection{Beyond the Magellanic Clouds}

Besides star clusters in the Magellanic Clouds, LSST will image star clusters in other Local Group galaxies.
At larger distances crowding will present even more substantial challenges than it does in the Magellanic Clouds, limiting our ability to study these star clusters using their resolved color magnitude diagrams.
Even if LSST only provides photometry of the brightest giants in the cluster outskirts, imaging with Rubin will still allow the structural parameters  as well as the integrated photometry to be measured (see \citealt{2014MNRAS.442.2165H, 2019MNRAS.484.1756M} for a demonstration of ground based cluster studies in the Local Group.).

A number of Local Group dwarf galaxies are known to host star cluster systems.
Within the LSST survey footprint, the Fornax dSph hosts a surprising number of GCs - 6 - for its stellar mass \citep{2021ApJ...923...77P}, the dwarf irregular galaxy NGC 6822 hosts eight old GCs \citep{2015MNRAS.452..320V} and a population of younger clusters \citep{2000AJ....120.3088C, 2004PASP..116..497K}. 
On the outskirts of the Local Group the WLM dIrr  shows a single old GC \citep{1999ApJ...521..577H}.
In addition, the Eridanus II dSph hosts a single low mass star cluster \citep{2015ApJ...805..130K, 2016ApJ...824L..14C} and the dIrr IC 1613 hosts a small number of low mass star clusters \citep{2000PASP..112..594W}.

Besides providing homogeneous measurements of known star clusters, LSST will allow for a systematic search for star clusters in Local Group dwarf galaxies. Together with star clusters in the Milky Way and the Magellanic Clouds, these cluster populations can be directly compared with those in more distant galaxies (see next section).

\section{Extragalactic Star Clusters}

\subsection{Motivation}
Beyond the Local Group, star clusters cannot be resolved into their constituent stars and must be studied using their integrated light.
There is a wealth of previous imaging studies of star clusters young and old from the ground and from space.
Surveys such as the SLUGGS survey \citep{2014ApJ...796...52B} of massive early-type galaxies and their globular cluster systems and the LEGUS survey of star forming galaxies \citep{2015AJ....149...51C} have targeted representative samples of galaxies.
Our understanding of old globular clusters is dominated by studies of Local Group galaxies and early-type galaxies in the Virgo and Fornax clusters \citep{2004ApJS..153..223C, 2007ApJS..169..213J}. Large studies of young star clusters with HST such as LEGUS \citep{2015AJ....149...51C} and PHANGS-HST \citep{2022ApJS..258...10L} do provide representative but not volume-limited samples of star forming galaxies in the local universe. 
Much of the work on globular clusters has focused on massive early-type galaxies since they have populous GC systems and their relatively simple morphologies and light distributions make studying their star cluster populations easier.
Work on other galaxy types has been fragmented and heterogeneous, often galaxy-by-galaxy, making comparisons difficult.

Most imaging studies have utilised imagers with fields-of-view smaller than the full extent of the star cluster systems, and this is particularly acute for studies with HST.
At the distance of the Virgo Cluster (16.5 Mpc, \citealt{2007ApJ...655..144M}), a single HST ACS pointing covers 16 by 16 kpc.
While this field of view can comfortably contain the GC system of a dwarf galaxy, the globular cluster systems of more massive galaxies extend beyond 100 kpc (AM-1 is 120 kpc from the centre of the MW; \citealt{
2008AJ....136.1407D}; PAndAS-48 lies 160 kpc in projection from M31; \citealt{2014MNRAS.442.2165H}) thus carrying out an accurate census of the GC systems of luminous ellipticals requires imaging that covers a radial distance of $>$120~kpc from the galaxy center; \citealt{2004AJ....127..302R, 2003AJ....125.1908D}).
Only the widest-field ground based imagers (e.g. MegaCAM on CFHT, the twin MegaCAM at Magellan, OmegaCam on VST, Hyper Suprime-Cam on Subaru, the Dark Energy Camera on CTIO, the One Degree Imager on WIYN) can image the entire GC system of a massive galaxy in a single pointing.
Ideally, one wants the imaging to extend beyond the GC system in order to (for example) accurately quantify the radial extent of the GC system, determine the extent of intragroup/cluster GCs (see e.g. \citealt {2022MNRAS.516.1320C}) and to more easily estimate the level of contamination from foreground or background objects in the GC candidate sample.
Due to its wide field nature, LSST will allow entire GC systems to be imaged. Furthermore, selection of GC candidates via photometry in six filters will yield cleaner GC candidate samples and while its high image quality will help eliminate non-GC extended sources such as background galaxies with similar colors.
The wide wavelength coverage provided by LSST will also allow the physical properties of extragalactic star clusters to be better constrainted.

 LSST presents the first opportunity to perform a large scale, homogeneous census for extragalactic globular clusters, as well as identifying intergalactic globular clusters and hypervelocity clusters \citep[e.g.,][]{2014ApJ...787L..11C}. We briefly discuss some of the impacts Rubin will have on our understanding of extragalactic star cluster systems including young star clusters, globular clusters and nuclear star clusters before discussing in more detail the science that will be enabled by the discovery of massive numbers of star clusters.


\subsubsection{Extragalactic Nuclear Star Clusters}
 Nuclear star clusters (NSCs) are dense star clusters at the centres of galaxies have masses ranging from $10^5-10^7$ $M_\odot$ and effective radii of only a few parsecs \citep{2020A&ARv..28....4N, 2017MNRAS.467.4180S}. 
Studies such as  \citet{2022ApJ...929...84B} have searched for massive black holes at the centers of nuclear star clusters, and the larger sample identified by Rubin will be able to extend these studies as well as provide full band photometric constraints on the nature of the star clusters.
\citet{2019ApJ...878...18S} studied NSCs in Virgo Cluster galaxies, finding a close connection between properties of NSCs and globular cluster systems (GCSs) of low mass galaxies. \citet{2021MNRAS.508..986Z} found a high occurrence of nuclear star clusters in faint Coma galaxies and that the mass of the host galaxy is the major driver of the efficiency of NSC formation and that the  environment plays a secondary role.
\citet{2022ApJ...927...44C} studied NSCs in the satellites of Milky Way-like galaxies in the Local Volume, finding lower nucleation fractions than for cluster galaxies, but similar connections between NSCs and GCSs.
Rubin studies will extend these studies to larger samples of galaxies over a greater range of environments and morphological types, providing additional constraints on formation models.

\subsubsection{Extragalactic Young Massive Clusters}

A significant fraction of stars forms in bound stellar clusters hence understanding clustered star formation is important for understanding star formation in galaxies.
Studies of the most massive young star clusters in the local universe can inform our understanding of how globular clusters formed.
Since the stars in a cluster are concentrated both spatially and temporally, the feedback from star clusters is likely more efficient than low density star formation.
Depending on how the stellar initial mass function is sampled, the most massive stars may only be found in young star clusters.

Young star clusters are typically found in star forming galaxies, which are morphologically complex. Due to this complexity, studying young star clusters in star forming disks typically requires space based imaging. The angular resolution of Rubin is $\sim 1$ arcsec, a factor of ten larger than the angular resolution of HST. 
For nearby galaxies (within $\sim 10$ Mpc), Rubin observations have similar spatial resolution to those of more distant galaxies observed with HST, for example the Hubble imaging Probe of Extreme Environments and Clusters (HiPEEC) survey of six merging galaxies \citep{2020MNRAS.499.3267A}.
The Rubin observations are likely limited to the fainter outskirts of galaxies and the brightest star clusters (i.e., the youngest and most massive).
The larger sample of galaxies provided by LSST will provide context to more detailed studies with HST and JWST \citep[e.g.][Adamo et al. in prep.]{2023ApJ...944L..14W}.

Active areas of research in young star clusters include how the initial cluster mass function varies between and within galaxies \citep[e.g.][]{2009A&A...494..539L, 2022ApJ...928...15W} and how the fraction of stars formed in bound clusters varies \citep[e.g][]{2000A&A...354..836L,2011A&A...529A..25S, 2010MNRAS.405..857G, 2015MNRAS.452..246A, 2017ApJ...849..128C, 2020MNRAS.499.3267A}.
Additionally, studies of the spatial correlations between young star clusters, molecular clouds and H II regions allow the time scales of each of these stars of star formation to be constrained \citep{2018MNRAS.479.1866K, 2023arXiv230503618P}.

Rubin studies of young star clusters will target the physics of star cluster formation, and the $u$-band photometry will be key to breaking the age-extinction degeneracy. 
In particular, the data will be able to target the following questions: 
\begin{itemize}
   
    \item How does the truncation mass of the initial cluster mass function vary with galaxy properties?
    \item How do the properties of nearby young star clusters compare with those of star forming clumps observed in high-redshift lensed galaxies \citep[e.g.,][]{mowla2022,Claeyssens2023}?
    \item What are the evolutionary links between young star clusters and older globular clusters?

\end{itemize}
\subsubsection{Extragalactic Globular Clusters}

Well-characterized extragalactic clusters can be used to extend our knowledge of galaxy structure, and the Rubin observations of extragalactic globular clusters will be able to measure ages, masses and metallicities from integrated light, probing different environments and galaxy ages. A further discussion of this takes place in Section \ref{section:inference}. 




\subsubsection{Science Cases}

\textbf{The dark matter content of galaxies:\ }
Globular clusters provide a number of probes of the dark matter content of galaxies, including the globular cluster-halo mass scaling relation, the timing problem in dwarf galaxies, as tracers of the galaxy potential and as tracers of the dark matter surface density, as is described in the next paragraph.

Recent literature has claimed that there is a one-to-one relationship between mass in globular clusters and galaxy halo mass over a wide range of halo masses \citep{1997AJ....114..482B, 2009MNRAS.392L...1S, 2014ApJ...787L...5H, 2018MNRAS.481.5592F, Zaritsky2022}.
These results imply that counting the size of a galaxy's GC population would be an observationally inexpensive way to measure its halo mass.
While there is strong evidence for the one-to-one relation in massive galaxies \citep[e.g.][]{2016MNRAS.458L..44F}, there are
theoretical reasons \citep{2019MNRAS.482.4528E, 2019MNRAS.488.5409C, 2020MNRAS.498.1050B} to expect that the relation breaks down in low mass galaxies,
In fact, in spite of some Local Group dwarf galaxies containing relatively high numbers of GC for their masses (see Section 3.3), there must be a breakdown at some point, since most of the lowest-mass galaxies do not host globular clusters \citep[e.g.][]{2022ApJ...926..162E}. 
The globular cluster mass-galaxy halo mass relation shows increased scatter at low galaxy masses \citep{2018MNRAS.481.5592F} so again, constraining it with larger numbers of galaxies will be beneficial.

It was first noted by \citet{1976ApJ...203..345T} that the half-dozen globular clusters in the Local Group Fornax dSph have ages much greater than their calculated orbital decay times.
Dwarf galaxies are presumed to be dark matter dominated.
If their central dark matter profiles are cuspy, as predicted by collisionless cold dark matter only simulations, one or more of the GCs in Fornax dSph should have inspiralled to the centre due to dynamical friction.
If their central dark matter profiles are cored, either due to baryon feedback or more exotic dark matter physics, the effects of dynamical friction are weaker, and less inspiralling would occur.
Globular cluster systems of these galaxies thus place important constraints on the nature of dark matter \citep{2021PhRvD.104d3021B}.
An issue with such studies is that any given galaxy has only a small number of clusters. One way around this is to `stack' cluster systems as in \cite{2022MNRAS.511.1860S}, and hence the larger number of galaxies observed by Rubin will be of enormous benefit.

There is a long history of using GCs as test particles in larger galaxies' halos.
Spectroscopy is required to obtain radial velocities for mass modelling, generally preceded by imaging to identify target GCs and to quantify their spatial distribution.
Simulations show that good kinematic information of at least 150 GCs per galaxy is required to recover the mass and radial distribution of the DM halo using dynamical models in extragalactic systems (\citealt{hughes2021}). 
Nonetheless, we can use the projected number counts of GC populations obtained with Rubin/LSST, alongside their galaxy stellar masses, to trace the structure of their host DM haloes, which is much less observationally demanding than observing the faint and diffuse stellar halo in the galactic outskirts \citep{reinacampos2021}.

\textbf{Low surface brightness science:\ }
Star clusters can make significant contributions to our understanding of the low surface brightness universe. 
Intragroup and intracluster GCs \citep[e.g.,][]{taylor2017,harris2020} can be used to probe kinematics of large-scale dark matter halos and trace galaxy assembly history.
They can be used for similar purposes on smaller spatial scales where found associated with stellar streams and shells in individual galaxy haloes \citep[e.g.,][]{Kang2022, Veljanoski2014}.
Star clusters in ultra-diffuse galaxies (UDGs) have recently received significant attention as both signposts to find UDGs and as indicators of UDG origins \citep[e.g.,][]{Li2022,Jones2023}
Rubin/LSST will provide a large sample of UDGs as well as `normal' dwarf galaxies allowing comparison between the GC populations of different types of dwarf galaxies.

\textbf{Star cluster formation:\ }
There are two broad models for the origin of GCs. 
Motivated by the old GCs of the MW and the lack of GC mass star clusters forming today in the MW the older model \citep[e.g.][]{1984ApJ...277..470P, 2015ApJ...808L..35T} of GC formation requires special conditions in the early Universe.
In the more modern model \citep[e.g.][]{1997ApJ...480..235E, 2010ApJ...712L.184E, 2015MNRAS.454.1658K}, motivated by observations of young star clusters with similar or higher masses to GCs in local starburst galaxies \citep[e.g.][]{1995AJ....109..960W, 1996AJ....112..416H, 2020MNRAS.499.3267A}, all star cluster formation shares a common process.
These two models make very different predictions about the age distribution of GCs.
In the older model all GCs should have similar old ages while in the modern model, the age distribution varies galaxy-to-galaxy depending on the formation history; since a high star formation rate density is required to form $>10^5$~\Msun\ of stars within a few parsecs in a few Myr, massive star clusters trace periods of high star formation.
Evidence for this modern model has emerged via observations of younger GCs in lower mass galaxies \citep[e.g.][]{
2006ApJ...646L.107C, 2006MNRAS.372.1259S, 2008A&A...489.1065M, 2014AJ....147...71P, 2020A&A...637A..27F} and from studies that find a connection between assembly history of galaxies and the ages and metallicities of their GCs \citep[e.g.]{2010MNRAS.404.1203F, 2019MNRAS.490..491U, 2020MNRAS.498.2472K}.

Simulations \citep[e.g.][]{2014ApJ...796...10L, 2018MNRAS.475.4309P, 2021MNRAS.505.5815V, 2023MNRAS.521..124R} based on these models make quantitative predictions that can be tested with observation from LSST.
The volume limited sample of GC systems provided by Rubin will be easier to compare with simulations than existing targeted studies.
Observations of young star clusters will be used to constrain models of star cluster formation.

\textbf{Galaxy formation:\ }

Assuming that GCs are the natural outcome of intense star formation, the properties of GCs can be used to study the formation and assembly history of their host galaxies.
The ages of GCs directly indicate when intense star formation occurred.
By inverting the redshift dependent galaxy mass-metallicity relation, the metallicity and age of a GC can be used to infer the likely mass of the galaxy it formed in \citep{2019MNRAS.486.3134K}.
Measurements from multiple GCs can be combined together to study the mass assembly history of a galaxy.
Since more massive and earlier mergers deliver their GCs on tighter orbits \citep{2020MNRAS.499.4863P}, by combining the ages, metallicities and orbits of GCs the assembly history of a galaxy can be reconstructed in detail as was done by \citet{2020MNRAS.498.2472K} for the MW.

Even when limited more information is available, \citep{2019MNRAS.486.3134K} showed that correlations exist between global GC properties like the number of GCs, the mean age and the slope of the age-metallicity relationship and the assembly history of a galaxy at fixed stellar mass.
LSST will provide a huge number of galaxies with well characterised GC systems.
Assembly histories derived from these GC systems can be compared to the predictions of simulation of galaxy formation.
LSST will also act as the input catalogue and provide context to spectroscopic studies that would enable more detailed assembly history studies.

\textbf{Globular cluster systems as distance indicators:\ } The GC luminosity function has long been promoted as a distance indicator, with systematic uncertainties at the 0.3~mag level \citep{2010ApJ...717..603V, 2012Ap&SS.341..195R}.
GCs are also a systematic in surface brightness fluctuations, where they need to be masked lest they contribute extra power to the fluctuations \citep{jensen2015}.
Mean GC sizes have also been proposed as standard rulers \citep{2010ApJ...715.1419M}. 
Although Rubin will only be able to measure reliable sizes for relatively nearby GCs, due to the nature of the survey, it will  increase the numbers of cluster systems sampled which will naturally allow the details of this distribution to be further probed.

\textbf{Stellar population variations in globular clusters across galaxies:\ } 
Over the past two decades many works have demonstrated non-linear color-metallicity and color-color relations in GC systems (e.g. \citealt{chies12}, \citealt{blakeslee2012}, \citealt{2019ApJS..240....2L}, \citealt{fahrion20}).
\citet{2012MNRAS.426.1475U, 2015MNRAS.446..369U} found that the relationship between GC color and metallicity varies between galaxies.
Recently, it was found that the color-color relation of GCs in different regions of the Virgo cluster behaves differently, implying that cluster formation happens differently in different environments (\citealt{powalka2016}).
There is also evidence for variation in the age distribution of globular clusters between galaxies \citep[][e.g.]{chies11b, 2019MNRAS.482.1275U}.
The range of galaxy environments and morphologies probed by these studies is incomplete: for example, they do not include clusters belonging to low surface brightness galaxies.
The homogeneous, ultra-deep multi-band imaging from LSST/Rubin will allow us to obtain a complete census of color-color relations up to $z=0.05$ and investigate this in unprecedented detail. Synergies with Euclid and Roman will extend the baseline to redder colors, as detailed in Section \ref{sec:synergies}.

\textbf{High-Energy Astrophysics:\ }
Young star clusters, particularly the more massive ones, may also provide clues to the formation channels for mass gap black holes \citep[e.g.][]{Mapelli14, rastello20, Rastello21, dicarlo21}. One of the best observational candidates for an intermediate mass black hole,  ESO 243-49 HLX-1 \citep{Farrell09}, may also be hosted by a young star cluster \citep{Farrell12}.  Rubin's unprecedented survey will provide a comprehensive data-set on clusters that will allow for multi-wavelength follow-up to provide further observational constraints on X-ray binary and intermediate mass black hole candidates in young massive star clusters \citep[e.g.][]{Rangelov12}. 

Globular clusters are also intimately linked with high-energy astrophysical events: novae \citep{2013ApJ...779...19K,2019ApJ...874...65D}, supernovae \citep{2012A&A...539A..77V,2013ApJ...762....1W}, fast radio bursts \citep{Kirsten2022} and many X-ray binaries, including ultraluminous X-ray sources (ULXs) \citep{maccarone07}. 
The remnant population of globular clusters can shine light on understanding the progenitors of transients, including gravitational wave events \citep{2020MNRAS.497..596D, 2022arXiv221016331A}.
Due to their relatively simple stellar populations, it is easier to measure the age and metallicity of a star cluster than field star light, allowing the progenitor of a transient to be constrained.
Unlike the stellar progenitors of transients, star clusters will survive transients relatively unchanged, allowing the progenitor of past transients to be studied.

Many of the ULXs in globular clusters have been identified using archival \textit{Hubble Space Telescope} and \textit{Chandra Observatory} observations of galaxies in the Virgo and Fornax clusters (e.g. \citealt{dage2021}). However, this method is observationally biased towards identifying clusters in the inner regions of the host galaxies;  \textit{Hubble} imaging cannot survey the entire cluster system of a massive galaxy. \textit{Chandra} has a much larger field of view, and the practical outcome is that detected X-ray sources in the galaxy outskirts are often left unprobed due to lack of wide-field optical coverage. 

X-ray population studies of extragalactic star clusters can also reveal the  X-ray binary populations, allowing for a model of the X-ray luminosity functions of star clusters \citep{2017ApJ...841...28P,2020ApJS..248...31L, 2023ApJ...947...31H}. These scaling relationships have further impacts on low-luminosity AGN identification, and feedback in AGN \citep{2022ApJ...936L..15C}. 
Rubin's more systematic and wide-field coverage will enable optical classification of the hosts of many of these sources and provide more spatially complete surveys of the LMXB content of extragalactic cluster populations.

\subsection{Tasks}
\subsubsection{Source detection and measurement}
Before sources can be classified as star clusters, they must be detected and their properties (flux and structural parameters) measured.
This process can be broken down into three regimes on the basis of background galaxy surface brightness and the morphological complexity.
For the faintest background galaxy light, such as the outer halos of galaxies and the intracluster light of galaxy clusters, the output of the default LSST source detection and photometry pipeline will likely suffice.
For brighter galaxies with relatively simple morphologies, i.e. early-type galaxies, the established approach is to subtract the galaxy light, either using a parametric model for the galaxy light or using a median filter image, perform the source detection step on the subtracted image then perform photometry on the original image.
For more morphologically complex galaxies (i.e. star forming galaxies), a more sophisticated deblending and source measurement process will be required.
Best seeing stacks would be quite useful for size measurements and source detection in crowded fields.
Understanding the completeness of source detection steps is critical, and comparison with HST and JWST imaging can be used to test source detection and measurement at higher spatial resolution.
However, most extragalactic star cluster studies with HST have utilized shallower imaging than Rubin will provide, thus new approaches will be needed for characterization of faint star clusters.

\subsubsection{Source classification}
Before star clusters can be studied they must be distinguished from foreground stars and background galaxies.
Fortunately, the properties of star clusters are intermediate between those of stars and galaxies, making selecting star clusters a more general case of star-galaxy separation \citep[e.g.][]{2012ApJ...760...15F, 
2020AJ....159...65S}.
Old, massive star clusters have regular, round morphologies.
Any objects with irregular or spiral morphologies, or that are highly elongated, can be classified as background galaxies. 
Only for the closest ($\sim 1$ Mpc) star clusters, where their diameters are larger than the FWHM PSF, can individual stars be seen in their outskirts.
In young low mass clusters, a handful of stars can dominate the light, giving a more irregular appearance.
At least for star clusters within a few tens of megaparsecs, star clusters are less extended than a significant fraction of background galaxies with the same apparent magnitude. 
Thus relatively clean samples of star clusters can be selected by only using size and magnitude cuts \citep[e.g.][]{2009ApJS..180...54J}.
Previous work \citep[e.g.][]{2009ApJ...699..254H} has shown that the effective diameter of a star cluster needs to be at least 10\% of the PSF FWHM for the diameter to be effectively measured.
For the median effective seeing of the survey (1.0 arcsec) and a typical GC with a radius 3 pc, this corresponds to a distance of 6 Mpc; for the best effective seeing expected (0.6 arcsec) this corresponds to a distance of 10 Mpc.
Thus, LSST will only be able to effectively separate stars and star clusters on the basis of their sizes within a few Mpc \citep[e.g.][]{2021ApJ...914...16H}.
However, even at larger distances, the high spatial resolution provided by Rubin will allow at least some of the background galaxy population to be separated from star clusters.

Since star clusters consist of stars with a range of temperatures, their position in color-color space is different than individual stars.
Compared to a star with same color in the blue, say $(u - r)$, a star cluster will have a redder color at longer wavelengths, say $(i - y)$.
Since star clusters formed their stars in a single burst and have evolved passively since, their position in color-color space is also different from actively star forming galaxies, with star clusters being bluer at longer wavelengths than galaxies with the same colors at shorter wavelengths.
Since the vast majority of star clusters that will be observed by Rubin are at redshift $z < 0.02$, the effects of redshift on the colors are minor compared to distant galaxies \citep{2014ApJS..210....4M}.

Finally, given that extragalactic star clusters are at distances of a megaparsec or more, their proper motions and parallaxes are insignificant, allowing anything with a non-zero proper motion or parallax to be selected as a star.
Gaia astronometry can be used for the nearest and most massive star clusters \citep[e.g.][]{2020ApJ...899..140V}, with Rubin astrometry used later in the survey for fainter sources.

As a classification problem, selection of star clusters is well suited to machine learning techniques.
Recent experiments with machine learning selection of clusters on either images \citep{2021ApJ...907..100P, 2022MNRAS.509.4094T} or photometric catalogs \citep{2022MNRAS.514..943B, 2022A&C....3900555M}  show promise.
Samples of star clusters, foreground stars and background galaxies from space based imaging (HST or JWST) can be used to train and validate classification techniques with Rubin data; other efforts (photometric redshift pipelines from the LSST Galaxies collaboration, star-galaxy separation techniques from SMWLV collaboration) may also prove useful.
Spectroscopically-confirmed clusters provide another possible training set (radial velocity measurements clearly select between foreground stars, star clusters and background galaxies) but pre-targeted spectroscopy suffers from selection biases.
Untargeted spectroscopy with integral field units (e.g. MUSE on the VLT) may prove useful to avoid such biases.

\subsubsection{Star cluster property inference}
\label{section:inference}
To learn about star clusters' formation and evolution and that of their parent galaxy, their physical properties need to be inferred.
Masses, ages and metallicities of stellar populations can be measured by fitting stellar population models to observed photometry.
Most imaging studies of GC systems have relied on imaging in two or three filters, often $gri$ or their analogues, preventing the effects of age and metallicity from being disentangled.
More filters and a wider wavelength coverage improve the constraints from fitting stellar population models, and the large number of filters provided by Rubin helps.
As can be seen in the color-color plots in Figure \ref{fig:lsst_colours}, while strong degeneracies exist between age, metallicity and extinction for $gri$, the addition of $uzy$ allows these degeneracies to be (partially) broken.
Priors on extinction values can be obtained from far-infrared observations and/or SED-fitting the integrated light of the host galaxy, with a complication being the lack of a way to tell whether a given star cluster is in front of, embedded in, or behind its host galaxy.
Spectroscopic observations will be useful for testing the ages and metallicities from Rubin photometry, with careful consideration of selection effects as noted above. 

\begin{figure*}
\begin{center}
     \includegraphics[width=512pt]{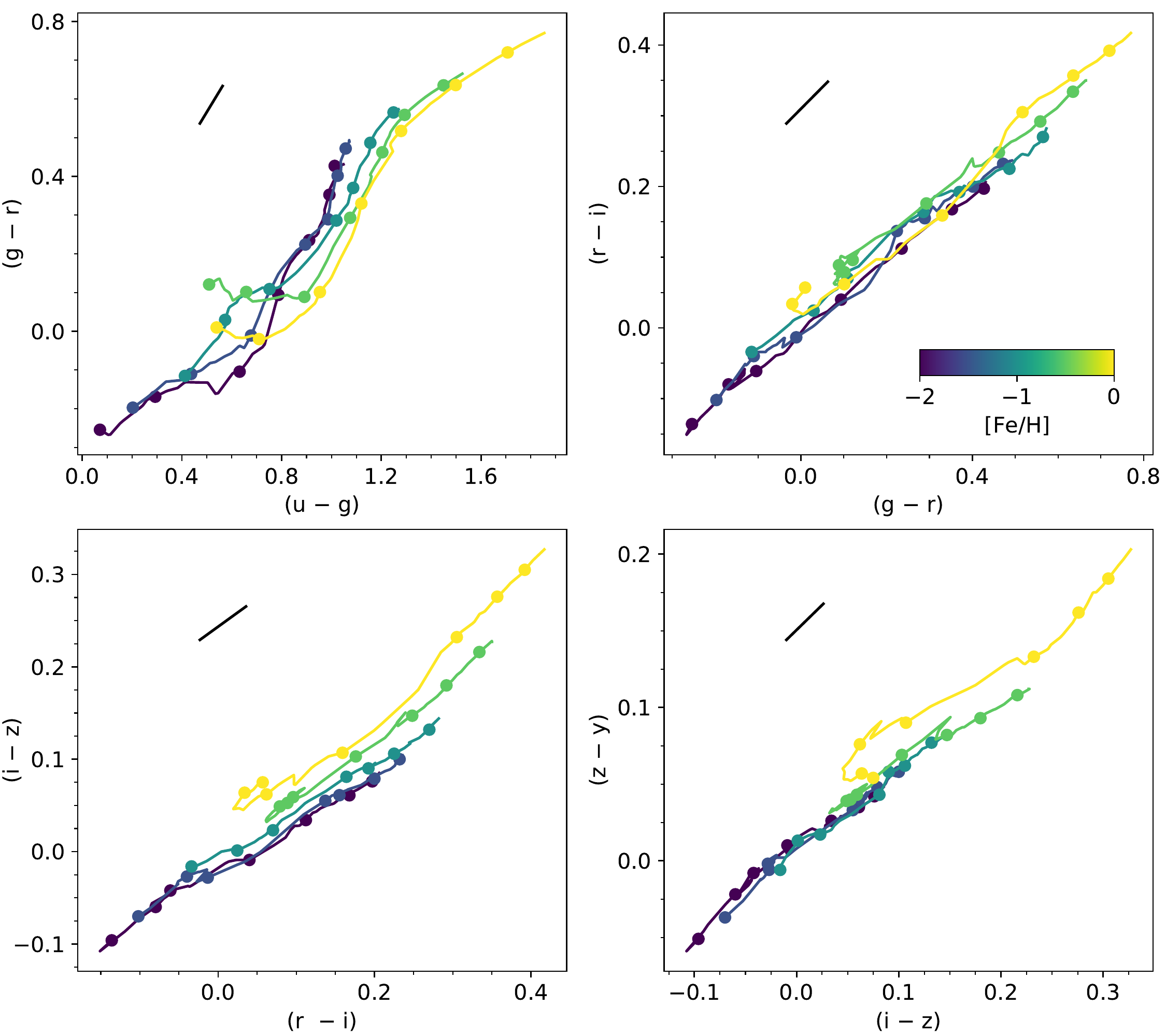}
     \caption{Color-color plots for model stellar populations in the Rubin bands calculated using \textsc{FSPS} \citep{2009ApJ...699..486C, 2010ApJ...712..833C}.
     In each panel the lines show model tracks from 100 Myr to 14.1 Gyr for five metallicities color coded from [Fe/H] $=-2$ in dark purple to [Fe/H] $= 0$ in yellow.
     The circles show the models at 100 Myr, 200 Myr, 500 Myr, 1 Gyr, 2 Gyr, 5 Gyr and 10 Gyr.
     The black line corresponds to the reddening vector for $E(B-V) = 0.1$.
     Although strong degeneracies exists between age, metallicity and extiction for $(g - r)$ versus $(r - i)$, the addition of $u$, $z$ and $y$ allows for these degeneracies to be broken.}
     \label{fig:lsst_colours}
\end{center}
\end{figure*}

Multi-band photometry of extragalactic star clusters gives ages, metallicities, and relative masses. 
To measure the mass and physical extent of an extragalactic star cluster, its distance is required. 
Most star clusters are too close for reliable photometric redshifts \citep{2014ApJS..210....4M}, and in any case peculiar velocities comprise a large part of the observed redshift at the expected distances.
The distance of an individual cluster is generally inferred from association with a host galaxy, which can be ambiguous if multiple galaxies are projected close together on the sky; for clusters in such locations, probabilistic galaxy membership better quantifies the limits of knowledge.
With a measured distance and angular size, physical size can also be measured. 
Physical sizes of old star clusters are of interest to studies of their dynamical evolution; physical sizes of younger clusters also constrain the star formation process.

\subsection{Deliverables}

The ideal deliverables from a Rubin study of extragalactic star clusters include both
a catalog of star clusters and a catalog of star cluster systems.
The catalog of star clusters would include photometric and structural/size measurements;
derived mass, age and metallicity; probabilistic classification (i.e. likelihood of being a star cluster; and probabilistic indication of parent galaxy).
The catalog of star cluster {\em systems} would include estimate of population size, spatial distributions, mass functions, and age and metallicity distributions, taking into account the various selection effects of the Rubin observations.






\subsection{Expected extragalactic star cluster sample}

To allow classification of sources as star clusters and to allow some inference on their properties requires that sources be detected in multiple bands.
For old stellar populations, typical of most GCs (using the predictions of the \textsc{FSPS} models of \citealt{2009ApJ...699..486C, 2010ApJ...712..833C}), and the expected depths of the final survey, a star cluster will first be detected in $r$ and $i$, then in $g$ and $z$, then in $y$ and finally in $u$, with more metal poor GCs being more easily detected in the blue and more metal rich in the red.
Since $(r - i)$ does not provide on its own a wide enough color baseline for effective selection or stellar population studies, we require detection in $g$ to be the minimum required to study an old star cluster. For young (less than a few hundred million years) star clusters, $u$-band photometry is important in disentangling the effects of age and extinction.





\begin{table}[]
\caption{\label{tab:gc_predict}Predictions for extragalactic star clusters observed with LSST.
(a) Rubin band.
(b) Apparent magnitude median single visit depth for the WFD based on the \texttt{baseline\_v3.0\_10yrs} OpSim run
(c) Apparent magnitude median coadd depth corrected for galactic extinction for the WFD footprint based on the \texttt{baseline\_v3.0\_10yrs} OpSim run.
(d) Total number of GCs observed in the entire LSST footprint over 10 yr based on the \texttt{baseline\_v3.0\_10yrs} OpSim run.
(e) absolute magnitude for a $2 \times 10^{5}$ M$_{\odot}$, 12.6 Gyr, [Fe/H] $= -2$ GC.
(f) Maximum distance in Mpc GC a GC with absolute magnitude from (e) will be seen given the depth given in (c).
(g) and (h) Same as (e) and (f) but for a [Fe/H] $= -1$ GC.
(i) and (j) Same as (e) and (f) but for a [Fe/H] $= 0$ GC.
(k) and (l) Same as (e) and (f) but for a $2 \times 10^{4}$ M$_{\odot}$, 100 Myr, [Fe/H] $= 0$ YMC.}
\centering
\begin{tabular}{c  c c  c  c c  c c  c c  c c}
(a) & (b) & (c) & (d) & (e) & (f) & (g) & (h) & (i) & (j) & (k) & (l) \\
Band & $m$ & $m$ & $N_{GC}$ & $M$ & $D$ & $M$ & $D$ & $M$ & $D$ & $M$ & $D$ \\
 & Visit & 10 yr & & \multicolumn{2}{ c }{[Fe/H] $= -2$} & \multicolumn{2}{ c }{[Fe/H] $= -1$}& \multicolumn{2}{ c }{[Fe/H] $= 0$} & \multicolumn{2}{ c }{100 Myr} \\
\hline
$u$ & 23.60 & 25.45 & $4 \times 10^{5}$ & $-6.40$ & 24 & $-5.86$ & 19 & $-4.56$ & 11 & $-7.50$ & 41 \\
$g$ & 24.41 & 26.54 & $1 \times 10^{7}$ & $-7.44$ & 64 & $-7.12$ & 55 & $-6.36$ & 39 & $-8.08$ & 86 \\
$r$ & 23.98 & 26.71 & $3 \times 10^{7}$ & $-7.86$ & 82 & $-7.69$ & 75 & $-7.11$ & 58 & $-8.13$ & 92 \\
$i$ & 23.39 & 26.20 & $2 \times 10^{7}$ & $-8.07$ & 72 & $-7.96$ & 68 & $-7.51$ & 56 & $-8.20$ & 76 \\
$z$ & 22.79 & 25.51 & $9 \times 10^{6}$ & $-8.15$ & 53 & $-8.10$ & 52 & $-7.83$ & 46 & $-8.30$ & 56 \\
$y$ & 22.00 & 24.73 & $3 \times 10^{6}$ & $-8.19$ & 38 & $-8.18$ & 38 & $-8.03$ & 35 & $-8.36$ & 41 \\
\end{tabular}
\end{table}

Based on the well-studied globular cluster luminosity function (GCLF, e.g. \citealt{jordan07}) that peaks at $M_g\sim-7.5$, one can estimate that LSST will reach the turnover magnitude (TOM) of GC systems at 40\,Mpc with single estimated exposures. For more nearby GC systems such as galaxies belonging to the Fornax Cluster, with one exposure, LSST will reach 1.5\,mag fainter than the TOM of the GCLF with a single exposure. 
After 10 years, we estimate that LSST will be able to reach the TOM of the GCLF of systems $\sim$100 Mpc away and 1.5 magnitudes brighter than the TOM for systems $\sim$200 Mpc (z$\sim$0.05).
Thus, LSST will be able to provide a complete census of GC systems up to z$\sim$0.05.

To estimate the number of GCs LSST will image, we combined the \citet{2022ApJ...926..162E} relationship between galaxy stellar mass and mass in GCs with the \citet{2022MNRAS.513..439D} galaxy stellar mass function.
By assuming an average GC mass of $2 \times 10^{5}$ M$_{\odot}$, we calculated a GC volume density of 7.6 Mpc$^{-3}$.
Using version 3.1 of the FSPS stellar population modelling code \citep{2009ApJ...699..486C, 2010ApJ...712..833C}, we calculated the absolute magnitudes of a GC with a stellar mass of $2 \times 10^{5}$ M$_{\odot}$, an age of 12.6 Gyr (corresponding to a formation redshift of $z = 5$) and a metallicity of [Fe/H] $= -1$ in each of the Rubin bands which we give in Table~\ref{tab:gc_predict}.
Assuming a Gaussian GC luminosity function centred on these magnitudes with dispersion of 1.2 mag (based on the MW's $V$-band GC luminosity function from \citealt{2001stcl.conf..223H}), by using the Healpix maps of coadded $5 \sigma$ depth corrected for Galactic extinction from the \texttt{baseline\_v3.0\_10yrs} OpSim  \citep{2014SPIE.9150E..15D} run, we can calculate the number of GCs across the sky LSST will likely detect over the course of the 10 year survey.
These estimates are provided in Table~\ref{tab:gc_predict}.
We also provide the median distances the model GC would be observed out to in the WFD footprint.
These numbers are optimistic given that they do not account for the lower limiting magnitude due to the higher background caused by the host galaxy light, for the effects of crowding, or for any extinction internal to the host galaxy.
We can use the same assumptions to estimate LSST will survey 100 000 galaxies each with at least 10 GCs detected in all of $griz$.

To account for the effects of metallicity we also calculated model photometry $2 \times 10^{5}$ M$_{\odot}$ 12.6 Gyr GCs with [Fe/H] $= -2$ and [Fe/H] $= 0$.
The metal poor GC is bluer and brighter while the metal rich GC is redder and fainter although we note that measurements of the GC mass-to-light ratio in the MW and M31 \citep[e.g.][]{2011AJ....142....8S, 2020MNRAS.492.3859D} show a constant $V$-band mass-to-light ratio with metallicity.
We provide the absolute magnitudes and median distance limits in Table~\ref{tab:gc_predict}.

The number of young star clusters LSST will observe is harder to estimate.
The fraction of stellar mass that forms in bound clusters and the truncation mass of the initial cluster mass function depend on the star formation surface density, such that more rapidly forming regions form more mass in clusters and form more massive clusters.

In Table~\ref{tab:gc_predict} we give absolute magnitudes and median distance limits for a $2 \times 10^{5}$, M$_{\odot}$ 100 Myr, [Fe/H] $= 0$ star cluster.
We expect that detecting young clusters should suffer more from the effects of the host galaxy background, crowding and extinction than GCs since they are typically found in and around active star forming regions.
The case of the extreme starburst Haro 11 is instructive.
Many of the young star clusters  in this galaxy studied by \citet{2010MNRAS.407..870A} with HST are bright enough to be detected in a single LSST visit. However, given the spatial resolution of LSST, at a distance of 85 Mpc the star clusters blur together into a handful of star cluster complexes.

\subsection{Synergies with Upcoming Missions}
\label{sec:synergies}
Rubin Observatory will have many synergies with upcoming and proposed missions. We highlight potential overlaps in star cluster science with a few of these.

Upcoming space-based wide field imaging missions like Euclid and the Roman Space Telescope will provide high resolution imaging and near-infrared photometry. The overlap between the Euclid Wide Survey and the extragalactic LSST WFD is about 9000 square degrees, or about half the LSST WFD and 60 \% of the Euclid survey. Rubin and Euclid have many synergies, specifically by using joint pixel deblended photometry on peripheral regions of Milky Way globular clusters, as well as dwarf satellites and nearby galaxies. Combining the two surveys will also allow for a volume-limited survey of globular cluster candidates \citep{2022zndo...5836022G}. About 4 million GCs will be detected at the 5 $\sigma$ level in both the Euclid VIS band and in the LSST $ri$ bands and about 800 000 GCs will be detected in all the Euclid bands as well as the $grizy$ LSST bands. About a quarter of the GCs detected in all four Euclid bands will be detected in $u$-band by LSST. Near infrared photometry would provide better star-globular cluster-galaxy separation  \citep[e.g.][]{2014ApJS..210....4M, 2018A&A...611A..21C} and provide stronger constraints on metallicity \citep[e.g.][]{2002A&A...391..441K, 2007ApJ...669..982C, chies12} and even age \citep[e.g.][]{chies11b, 2012MNRAS.420.1317G, 
2015A&A...581A..84T}. 
High spatial resolution imaging from space would allow star, globular cluster, and galaxy separation out to much greater distances using angular sizes.

The Nancy Grace Roman Telescope's high precision astrometry and wide-field of view will enable studies of the entire tidal radius of a globular cluster%
\footnote{\url{https://www.stsci.edu/files/live/sites/www/files/home/roman/_documents/roman-capabilities-stars.pdf}}, as well as studies of multiple stellar populations in globular clusters and stellar kinematics of the cluster \citep{2019arXiv190305085B}.
The field-of-view Roman's Wide Field Instrument is also well matched to the size of GC systems nearby galaxies (140 by 70 kpc at a distance of 10 Mpc) allowing for high resolution near-infrared imaging of a galaxies GC system in a single pointing.
Roman's High Latitude Wide Area Survey will likely provide deeper (25.8 to 26.7 AB, similar depths to LSST) NIR photometry than Euclid over a smaller area ($\sim 1700$ square degrees \url{https://roman.gsfc.nasa.gov/high_latitude_wide_area_survey.html}).

Future planned and proposed UV and optical missions such as CASTOR \citep{2019clrp.2020...18C}, UVEX, STAR-X, CSST \citep[e.g.][]{2023MNRAS.tmp.1351Q} and ULTRASAT%
\footnote{\url{https://www.weizmann.ac.il/ultrasat/}} will provide valuable UV coverage to Rubin's full optical coverage. In particular, UV photometry would yield improved constraints on age, metallicity and extinction.

\subsection{Summary}
The Rubin dataset will be transformative for a broad range of star cluster science. 
Rubin observations will detect the lowest-mass stars in the nearest Milky Way clusters and the integrated light from the brightest clusters in galaxies at distances of 100~Mpc.
With these data, questions of formation and evolution of stars, star clusters, and galaxies can all be addressed.
This roadmap does not represent an exhaustive catalog of every topic that can be addressed with Rubin data: 
the buildup of galaxy haloes from dissolving star clusters and their relationship with stellar streams; 
and
stellar variability in Galactic clusters
are just a few examples of star cluster topics not mentioned above.
The most exciting prospect is the new questions that will be generated by this wealth of data: the long history of using star clusters to understand stars and galaxies is far from complete.

\begin{acknowledgments}
The authors thank the referee for their useful suggestions, and Marina Kounkel for helpful discussion.
CU acknowledges the support of the Swedish Research Council, Vetenskapsr{\aa}det. KCD acknowledges fellowship funding from Fonds de Recherche du Qu\'ebec $-$ Nature et Technologies, Bourses de recherche postdoctorale B3X no. 319864. LG acknowledges funding by an INAF Theory Grant 2022. 
PB acknowledges funding from an NSERC Discovery Grant.		ACS acknowledges funding from CNPq-314301/2021-6.		WIC gratefully acknowledges support from the Preparing for Astrophysics with LSST Program, funded by the Heising Simons Foundation through grant 2021-2975, and administered by Las Cumbres Observatory.	
MG gratefully acknowledges support by ANID/Fondecyt Project 1220724.  AP acknowledges support from CNPq (Conselho Nacional de Desenvolvimento Científico e Tecnológico) Brazilian agency. RAS gratefully acknowledges support from the National Science Foundation under grant number 2206828.
RSz acknowledges the support by the Lend\"ulet Program  of the Hungarian Academy of Sciences, project No. LP2018-7/2022.	
LV is supported by the National Aeronautics and Space Administration (NASA) under grant No. 80NSSC21K0633 issued through the NNH20ZDA001N Astrophysics Data Analysis Program (ADAP).		 MRC gratefully acknowledges the Canadian Institute for Theoretical Astrophysics (CITA) National Fellowship for partial support; this work was supported by the Natural Sciences and Engineering Research Council of Canada (NSERC).	EAH acknoweldges support from CAPES and CNPq.
\end{acknowledgments}
\bibliography{clusters}
\bibliographystyle{aasjournal}

\end{document}